\documentclass{article}

\usepackage[a4paper]{geometry}
\usepackage{amssymb}
\usepackage{amsmath}
\usepackage{mathtools}
\usepackage[sectionbib]{natbib}
\usepackage{graphicx}
\usepackage{float}
\usepackage{graphicx}
\usepackage{subcaption} 
\usepackage{tabularx}
\usepackage{booktabs}      
\usepackage{ragged2e}    
\usepackage{makecell}     
\usepackage{seqsplit}    
\usepackage{tikz}

\usepackage{tabularx,booktabs,ragged2e,adjustbox,caption}
\usepackage{longtable}
\usepackage{eurosym}
\usepackage{amsthm}
\usepackage{caption}
\usepackage{comment}
\usepackage[utf8]{inputenc} 
\usepackage[T1]{fontenc}    
\usepackage{hyperref}       
\usepackage{url}            
\usepackage{booktabs}       
\usepackage{amsfonts}       
\usepackage{nicefrac}       
\usepackage{cleveref}
\usepackage[font=small,skip=10pt,labelfont=bf,tableposition=top]{caption}
\usepackage[font=small,skip=0.3pt]{subcaption}
\usepackage[section]{placeins}
\usepackage{array}
\usepackage{xcolor}
\newcount\Comments  
\usepackage{color}
\definecolor{darkgreen}{rgb}{0,0.5,0}
\definecolor{purple}{rgb}{1,0,1}
\newcommand{\kibitz}[2]{\ifnum\Comments=1\textcolor{#1}{#2}\fi}

\newcommand{\secreview}[1]{#1}  

\usepackage{booktabs}

\usepackage{multirow}
\makeatletter
\newcommand\thefontsize{The current font size is: \f@size pt}
\makeatother

\usepackage{setspace}

\DeclareCaptionLabelFormat{andtable}{#1~#2  \&  \tablename~\thetable}

\title{StableAML: Machine Learning for Behavioral Wallet Detection in Stablecoin Anti-Money Laundering on Ethereum}

\author{
    \begin{tabular}{ccc}
        Luciano Juvinski$^{a}$ & Haochen Li$^{a}$ & Alessio Brini$^{a}$\\
        \texttt{luciano.silva@duke.edu} & \texttt{haochen.li2@duke.edu} & \texttt{alessio.brini@duke.edu}\\
    \end{tabular}
}

\date{}

\begin{document}

\maketitle
\vspace{3cm}
\thispagestyle{empty}

\begin{center}
    $^a$\textit{Duke University Pratt School of Engineering, 305 Teer Engineering Building, Box 90271, Durham, NC 27708, USA} \\
\end{center}

\textbf{Keywords}: Anti-Money Laundering, Stablecoins, Blockchain Analytics, Know Your Transaction, Machine Learning.

\begin{abstract}

Global illicit fund flows exceed an estimated \$3.1 trillion annually, with stablecoins emerging as a preferred laundering medium due to their liquidity. While decentralized protocols increasingly adopt zero-knowledge proofs to obfuscate transaction graphs, centralized stablecoins remain critical ``transparent choke points'' for compliance. Leveraging this persistent visibility, this study analyzes an Ethereum dataset to establish an empirical baseline for behavioral AML detection. Our findings demonstrate that domain-informed tree ensemble models achieve higher Macro-F1 score, significantly outperforming graph neural networks, which struggle with the increasing fragmentation of transaction networks. The model's interpretability goes beyond binary detection, successfully dissecting distinct typologies: it differentiates the complex, high-velocity dispersion of cybercrime syndicates from the constrained, static footprints left by sanctioned entities. This methodological approach provides actionable insights that align with industry shifts toward deterministic verification, informing the auditability and compliance requirements under regulations such as the EU’s MiCA and the U.S. GENIUS Act while minimizing unjustified asset freezes. By providing a high-precision behavioral classification of suspicious wallets, this approach contributes to raising the economic cost of financial misconduct while informing compliance practice under emerging stablecoin regulations.  
\end{abstract}
\vspace{3cm}

\section{Introduction}
Most forms of organized criminal activity have a financial component. Whether arising from corruption, trafficking, cyber-enabled fraud, or terrorist financing, illicit proceeds must be concealed, moved, and eventually reintegrated into the legitimate economy. This dependence on financial flows makes money laundering the cornerstone of criminal enterprises. In 2024, an estimated \$3.1 trillion in illicit funds flowed through the global financial system, fueling a shadow economy outside regulatory control \citep{verafin2024global}. 

Blockchain money laundering mirrors the traditional placement, layering, and integration stages of financial crime but occurs at a significantly higher speed. An estimated \$51 billion was laundered through cryptocurrencies in 2024, representing approximately 0.14\% of all on-chain transactions\footnote{\url{https://go.chainalysis.com/2025-Crypto-Crime-Report}}, which totaled \$9 trillion. Crypto-based laundering typically involves injecting illicit funds, obscuring their origins through complex and multi-hop transaction paths\footnote{A multi-hop transaction refers to the practice of routing funds through a sequence of intermediate wallet addresses before reaching a final destination, rather than a single direct transfer. In anti-money laundering contexts, this technique is commonly used to create a complex chain of custody, effectively distancing the funds from their illicit origin and complicating audit trails.}, and ultimately reintegrating them into the legitimate economy. This environment exhibits a dual nature: while public ledgers provide technical transparency by making transaction histories visible, they simultaneously preserve pseudonymity by linking activity only to cryptographic addresses. Paradoxically, rather than preventing illicit practices, this open and permissionless architecture has enabled increasingly sophisticated obfuscation techniques involving decentralized exchanges, token swaps, and cross-chain bridges \citep{lin2023understanding,griffin2025cryptoaml}. Within this ecosystem, stablecoins have emerged as central vehicles for value transfer. Stablecoins are cryptocurrencies engineered to maintain a stable value relative to an external reference, such as a fiat currency or commodity, typically through collateralization or algorithmic mechanisms. By reducing price volatility, they function as media of exchange and units of account within both centralized and decentralized financial infrastructures \citep{clark2019sok}. Owing to these properties, stablecoins have become the dominant vehicle for illicit value transfer, accounting for over 84\% of all verified crypto fraud volumes in 2025 \footnote{\url{https://www.trmlabs.com/reports-and-whitepapers/2026-crypto-crime-report}}, driven by their high liquidity, interoperability, and deep integration across exchanges, decentralized finance (DeFi) protocols, and payment rails.

\begin{figure}[H]
    \centering
    \includegraphics[width=1.01\linewidth]{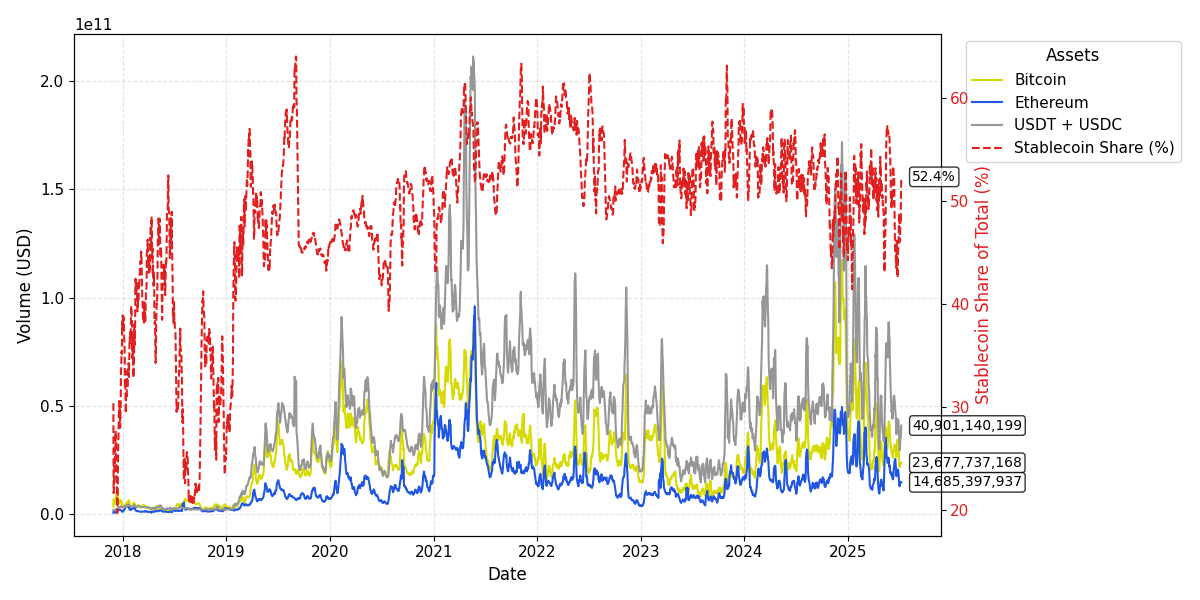}
    \caption{Historical transaction volumes for Bitcoin, Ethereum, and combined stablecoins.} 
    \label{fig:stablevscrypto}
\end{figure}

The Ethereum blockchain plays a particularly important role due to its dominance in Web3\footnote{Web3 refers to a paradigm for web-based applications that leverages blockchain technology, smart contracts, and token-based economic mechanisms to enable decentralized execution and coordination. In contrast to Web 2.0 platforms, which typically rely on centralized intermediaries, Web3 applications allow users to interact through decentralized applications (dApps) in which control over digital assets is mediated by cryptographic keys rather than centralized authorities.} activity and its high incidence of security events. Tether (USDT) and Circle USD Coin (USDC) together processed over \$10 trillion in transfer volume in 2025\footnote{\url{https://visaonchainanalytics.com/transactions}}, exceeding the settlement activity of both Bitcoin and Ethereum combined (see Figure~\ref{fig:stablevscrypto}). Although these assets support legitimate financial activity, their rapid settlement, which often allows transactions to finalize within seconds rather than days, combined with borderless accessibility and inherent pseudonymity, makes them attractive for money laundering \citep{pocher2023detecting}. Recent empirical findings further show that stablecoins are now the preferred medium for laundering and sanctions evasion, particularly following enforcement actions against obfuscation services such as mixer\footnote{Mixers are services designed to obscure transaction provenance by pooling funds from multiple users and redistributing them in a way that breaks the link between sender and recipient.} protocols \citep{griffin2025cryptoaml}.

Concurrent with these trends, stablecoin regulation is advancing rapidly across major jurisdictions. The EU’s Markets in Crypto-Assets (MiCA) Regulation (2023) and the U.S. GENIUS Act (2025) mandate that stablecoin issuers maintain full reserve backing, undergo independent audits, and comply with comprehensive Anti-Money Laundering (AML) requirements, while several Asian regulators have introduced comparable regimes. Among those, Singapore strengthened its oversight through the Payment Services Act and the 2023–2024 Stablecoin Regulatory Framework, while Hong Kong implemented its Stablecoin Issuer Regulatory Regime through recent amendments to the Anti-Money Laundering and Counter-Terrorist Financing Ordinance. China also regulates private stablecoin activity under the People's Bank of China Digital Currency Framework linked to its ongoing digital yuan initiative. 

In parallel with these regulatory efforts, major stablecoin issuers, such as Tether and Circle, have adopted wallet freezes, an enforcement mechanism in which the issuer leverages smart-contract privileges to prevent designated addresses from transferring, redeeming, or otherwise moving their tokens. Recent evidence shows that such freezes have immobilized over \$4 billion in USDT and more than \$1 billion in USDC, substantially increasing both the cost and the traceability of illicit transactions \citep{griffin2025cryptoaml}. These freezes trap criminal funds on-chain. They also prompt offenders to add intermediary hops, routing funds through a sequence of additional wallets and services, to circumvent restrictions. This often results in longer and less traceable transaction paths. Collectively, these regulatory and enforcement developments underscore the growing effectiveness of policy-driven interventions. However, while these measures serve to raise the economic cost of financial crime, they simultaneously drive illicit actors toward increasingly sophisticated evasion strategies to bypass them. This adversarial evolution creates a critical need for scalable detection frameworks that continuously monitor and adapt to shifting laundering behaviors on the blockchain. 

The core challenge is differentiating legitimate activities from illicit ones in high-throughput environments that process millions of transactions daily. Conventional rule-based systems, widely used in traditional finance, rely on static thresholds and deterministic heuristics such as flagging transfers above fixed monetary limits, detecting rapid bursts of activity, or blocking wallets previously labeled as high-risk. Although intuitive, these systems represent a staggering inefficiency, producing false positive rates that exceed 95\% \citep{liu2023graph}. This operational failure imposes massive economic overheads, estimated at \$136.5 billion in annual compliance costs in Europe alone, yet intercepts only 0.1\% of global criminal funds \citep{akartuna2022preventing}. Such inefficiencies create substantial burdens while failing to capture the adaptive behaviors characteristic of modern blockchain-based laundering.

This study addresses these limitations through three primary contributions. First, we construct \textit{StableAML}, a large-scale, domain-specific dataset focused exclusively on stablecoin flows. We collect suspicious and malicious Ethereum wallets identified through Etherscan reports, disclosures from partner blockchain security firms (e.g., SlowMist, PeckShield), and official sanctions and enforcement lists, specifically the U.S. Office of Foreign Assets Control (OFAC) Specially Designated Nationals list. We integrate these labels with the complete transfer history of USDT and USDC on Ethereum and process the data through a feature engineering pipeline that constructs 68 domain-specific attributes organized into four distinct categories: Interaction Features, Derived Network Features, Transfer-based Features, and Temporal and Direct Features. To the best of our knowledge, this is the first study to construct a large-scale labeled dataset exclusively focused on stablecoins. This specialized scope allows us to isolate unique token mechanics, specifically smart contract freezes and blacklist interactions, that are absent in native-asset models yet critical for identifying stablecoin-specific laundering typologies. While the predicate offenses underlying flagged wallets are heterogeneous, ranging from DeFi exploits and phishing operations to OFAC sanctions and issuer-level freezes, the subsequent act of moving, stabilizing, and exiting illicit proceeds tends to converge on a common set of stablecoin rails. Our study therefore targets the layering and integration phases of money laundering, in which these otherwise dissimilar categories share a common behavioral footprint. Moreover, in this emerging paradigm, the global visibility of the native chain will likely be obfuscated by privacy layers. Consequently, AML frameworks must evolve to operate within deterministic \textit{compliance tunnels}, where decision-making is restricted to exclusive, permissioned transfer logs. This validates a methodological pivot toward models that can extract robust risk signals solely from stablecoin contract events, as these regulated liabilities will serve as the persistent, auditable ``choke points'' in an otherwise private ledger economy. Emerging compliance instruments mandated under MiCA and the GENIUS Act, such as audit keys and proof-of-innocence schemes, reinforce this trajectory by requiring stablecoin issuers and regulated counterparties to verify the licit provenance of funds without exposing the underlying transaction graph.  

Second, we develop a comprehensive evaluation framework leveraging supervised machine learning to classify wallet behaviors, explicitly addressing the unique obfuscation techniques of the stablecoin ecosystem. Detecting illicit activity on-chain involves challenges distinct from traditional fraud: malicious actors rapidly dissipate illicit funds across vast wallet networks or utilize chain hopping strategies, routing funds through cross-chain bridges and decentralized exchanges, to fragment the transaction graph. Furthermore, emerging privacy technologies, specifically zero-knowledge–based shielding protocols (e.g., Railgun) and mixer services, increasingly obscure transaction linkages by cryptographically breaking the on-chain traceability between senders and recipients \citep{europol2020internet,force2021updated,manning2025futurecrime}. These factors create significant sparsity and structural disconnects in the transaction network. In this study, we deliberately evaluate Graph Neural Networks (GNNs) under these sparse conditions to simulate a \textit{Compliance Tunnel} scenario. As native transaction graphs become increasingly obscured by privacy-preserving protocols, stablecoin ledgers serve as the primary auditable channels. We test the models to determine their performance when restricted to these fragmented transfer logs.


Third, we enhance the interpretability of AML detection to support policy and compliance efforts. While this research focuses on the static classification of behavioral signatures, it provides the foundational analytical step necessary to bridge the gap between machine learning performance and regulatory interpretability, serving as a precursor to future real-time monitoring systems.
In this context, it is essential to distinguish between AML behavioral risk scoring and deterministic money laundering identification. While our dataset relies on verifiable malicious activity (e.g., OFAC sanctions, known DeFi hacks), strict legal identification of money laundering requires off-chain context and proof of predicate offenses. Therefore, our framework functions as an advanced triaging mechanism for compliance officers, scoring the behavioral risk of on-chain entities through obfuscation signatures, rather than providing absolute legal identification.

The remainder of the paper is organized as follows. Section~\ref{Sec:literature} reviews related work on blockchain-based AML detection, privacy challenges, and regulatory enforcement. Section~\ref{Sec:dataset} describes the dataset construction and feature engineering process. Section~\ref{Sec:models} outlines the modeling methodology, including the evaluated machine learning and graph-based approaches. Section~\ref{Sec:results} presents the experimental results and discusses their implications, including feature importance, detected activity typologies, and binary classification analyses. Finally, Section~\ref{Sec:conclusion} concludes the paper and highlights directions for future research.

\section{Literature Review}\label{Sec:literature}
Machine learning has long played an important role in fraud detection within traditional financial systems. Banks and payment networks routinely employ supervised models to identify credit card fraud and identity theft by learning patterns in user behavior, merchant characteristics, and transaction timing \citep{chen2018machine}.

In this traditional context, tree ensembles have demonstrated detection capabilities comparable to deep neural networks. Recent benchmarks report F1 scores exceeding 0.90 for tabular transaction data, reinforcing their viability without the computational overhead of complex graph-based approaches \citep{Gorte2023}.

Building on this foundation, the cryptocurrency literature applies machine learning and graph analysis to identify money laundering patterns in digital assets. \cite{shojaeinasab2024decoding} studies Bitcoin mixing and trains GNNs to classify links to mixers. Similarly, \cite{pocher2023detecting} detect illicit activity on the Bitcoin network using the Elliptic dataset to classify transactions as illicit or legitimate. The Elliptic benchmark itself was introduced by \cite{weber2019anti}, who first demonstrated the feasibility of graph convolutional networks for transaction-level AML on Bitcoin and provided the labeled subgraph that has since become the standard reference for this line of work. More recent extensions to Ethereum-specific AML detection \citep{lo2022inspection} confirm such architectural template, while highlighting the additional challenges introduced by the smart-contract layer.

In the broader Web3 context, \cite{lin2023understanding} introduce the EthereumHeist dataset to analyze addresses linked to hacks and scams. Their work demonstrates that programmable obfuscation techniques, specifically token swaps and cross-chain transfers, have become central mechanisms for concealing illicit fund origins.

From a methodological perspective, \cite{liu2023graph} introduce GTN2vec to identify laundering activity on Ethereum. While effective for native Ether transactions, their graph-learning approach relies on continuous transaction chains. Moreover, while advanced temporal architectures such as EvolveGCN \citep{pareja2020evolvegcn} and the Temporal Graph Network framework \citep{rossi2020temporal} dynamically update graph embeddings, these message-passing models fundamentally depend on persistent topological connectivity. This leaves a specific gap in addressing stablecoin ecosystems, where token mechanics and obfuscation intentionally fragment the transaction graph, rendering both static and dynamic GNNs highly sensitive to network sparsity.  

Looking beyond current graph limitations, \cite{berentsen2023introduction} outline a future landscape where privacy-preserving technologies, specifically ZKPs, fundamentally alter the availability of public transaction data. The widespread integration of cryptographic obfuscation, particularly Zero-Knowledge Proofs (ZKPs), significantly enhances data protection but deliberately breaks the continuous graph connectivity historically relied upon for fraud detection \citep{xue2022, solunke2024}. They argue that the economic demand for privacy will drive a shift toward \textit{selective transparency}, where transaction validity is decoupled from the public exposure of sensitive data. Complementing this view, \cite{chang2020blockchainimpact} examine blockchain adoption in financial services using qualitative evidence to highlight these monitoring challenges.

Extending the scope to stablecoins, \cite{vatsa2025stablecoin} provides a large-scale empirical analysis highlighting strong market concentration. The authors examine how regulatory initiatives, including the EU’s MiCA and the U.S. GENIUS Act, promote transparency but may also reinforce market dominance by imposing compliance costs that act as barriers to entry. Intersecting with this dynamic, \cite{fajri2026} highlight a growing regulatory paradox between AI-driven AML enforcement and stringent data protection laws. As privacy frameworks increasingly limit global on-chain visibility, monitoring must shift toward isolated, deterministic \textit{compliance tunnels}, such as stablecoin ledgers, to balance traceability with privacy rights. Complementing such a regulatory perspective, \cite{griffin2025cryptoaml} trace illicit flows to assess the effects of interventions like OFAC sanctions and asset freezes. Their findings show that while enforcement reduces illicit mixer volume, criminal actors migrate to DeFi swaps and cross-chain bridges. \cite{akartuna2022preventing} argue against ``one-size-fits-all'' prevention strategies. They emphasize that distinct asset classes like stablecoins require tailored detection models capable of addressing their unique risk profiles and operational contexts.

Collectively, these studies show that AML strategies must integrate advanced analytical models with evolving regulatory landscapes. However, existing research remains largely focused on native cryptoassets or dependent on continuous graph topologies. This creates a critical gap for supervised, feature-centric frameworks specifically tailored to the stablecoin ecosystem, where unique token mechanics require specialized detection strategies distinct from traditional graph-based tracing.

\section{The \textit{StableAML} Dataset}\label{Sec:dataset}

The Ethereum blockchain records billions of transactions. Our study narrows this scope to the value flow of the two dominant stablecoins, USDT and USDC. To isolate these movements, we query the ledger specifically for \texttt{Transfer} events emitted by the official smart contracts\footnote{Smart contracts are self-executing programs stored on a blockchain that run when predetermined conditions are met. In the context of stablecoins, these contracts autonomously manage the issuance, transfer, and balance tracking of tokens without a centralized intermediary.} between 2017--11--28 and 2025--08--08. Unlike raw transactions, which often involve complex contract interactions without actual value exchange, event logs allow us to capture definitive token movements.

\begin{table}[H]
\centering
\begin{tabular}{l l}
\hline
\textbf{Field} & \textbf{Value} \\
\hline
Transaction   & \texttt{\small 0xf8163c3d5ba77186ad4c6c93c4f3f92a88adb1b55cd0da34de2a44d33d4e20bd} \\
Sender Address   & \texttt{0x654Fae4aa229d104CAbead47e56703f58b174bE4} \\
Recipient Address & \texttt{0x000000000035B5e5ad9019092C665357240f594e} \\
Amount           & 1,092,761.61 USDT \\
Timestamp        & 2024-01-31 11:59:59 \\
\hline
\end{tabular}
\caption{Example of a real token transfer event extracted from the dataset. The transaction occurred on 2024-01-31 at 11:59:59~UTC and involved a transfer of 1,092,761.61~USDT.}

\label{table:transfer_example}
\end{table}

Transfer events remain visible even when privacy-enhancing technologies, which obscure transaction-level details, are employed. This choice resolves ambiguities inherent in raw transaction data, particularly in meta-transaction or relayer scenarios.

For example, consider a case where a Relayer (Wallet A) submits a transaction to execute a transfer on behalf of a User (Wallet B) to a Recipient (Wallet C). The blockchain's raw transaction ledger records Wallet A as the initiator, obscuring the true economic actor. However, the emitted \texttt{Transfer} event ignores the intermediary and correctly logs the token transfer from B to C.

By relying on these event-level records, we reconstruct value flows directly between counterparties and aggregate the observed transfers at the wallet level, constructing behavioral profiles for each address exploiting the transparency standards mandated by their technical design \citep{force2021updated,europol2020internet}.

The observational baseline for this study was constructed from the complete historical record of USDT and USDC transfers, encompassing 334,754,008 transactions and 54,998,597 unique wallet addresses. From this macro-population, 363,887 addresses were identified as malicious or suspicious. Crucially, all behavioral and topological features (e.g., \textit{2ndWithFlagged}, \textit{receiveMultipleSameValue}) were computed at the macro-network level prior to any sampling. We instantiated the complete 54.9-million-node transaction graph to calculate precise multi-hop metrics for every candidate node. This order of operations ensures that the topological features accurately reflect each wallet's true structural position within the complete Ethereum ecosystem, rather than being an artifact of a sparse subgraph.

Once this global feature extraction phase was completed, we applied a targeted filtering funnel to transition from the raw population to the labeled modeling dataset. To address potential label noise and ensure the robustness of the ``Normal'' class, we implemented strict inclusion criteria. First, we applied a minimum activity heuristic, excluding dormant or near-zero transaction addresses. This step prevents evaluation metrics from being inflated by trivial cases, ensuring the model focuses on distinguishing active legitimate users from laundering operations. Second, we applied a Proximity Exclusion Filter to the remaining benign candidates. This filter eliminates any otherwise unflagged address that falls within a 2- to 3-hop transaction radius of a known illicit node, ensuring they are not undetected accomplices and providing a structurally verified baseline for the Normal category.

Conversely, inclusion in the illicit categories (Blocklisted and Cybercrime) was governed by ground-truth certainty rather than activity thresholds. Because money laundering often relies on single-use or high-churn wallets (e.g., disposable hop-addresses), excluding low-activity nodes from these classes would erase the structural fingerprint of illicit layering. Therefore, illicit labels were strictly gated by deterministic on-chain events (e.g., decoding smart contract blacklist parameters) and corroborated intelligence.

The final 16{,}433 labeled wallets that constitute the modeling dataset are then drawn from this filtered candidate pool through stratified random sampling on the class label. Specifically, we anchor the sample on the \textit{Cybercrime} class, which is the smallest of the three with confirmed stablecoin activity, and draw \textit{Blocklisted} and \textit{Normal} addresses from the pool to reach a managed target distribution that supports discriminative learning. Because the two illicit categories together account for less than 1\% of active addresses at the macro-network level, a real-world-proportional sample would leave the model unable to learn the multi-hop layering signatures we aim to capture. Sampling on actual on-chain addresses, rather than synthetic minority oversampling such as SMOTE, preserves authentic behavioral and topological signal in every retained wallet. The resulting class proportions are reported at the close of the next subsection, and Section~\ref{Sec:models} describes the 80/20 partition into training and held-out test sets used for model fitting and final evaluation.

\subsection{Classification Groups}

We divide wallet addresses into three distinct classes to perform supervised learning and capture behavioral and regulatory heterogeneity within the stablecoin ecosystem. For reproducibility, the labels in each category are obtained from publicly accessible sources rather than from proprietary lookups, as described below:

\begin{itemize}
    \item \textit{\textbf{Normal}}: addresses with no negative flags
    from market-leading intelligence providers, specifically Chainalysis, TRM Labs, and Crystal Intelligence\footnote{\secreview{These providers publicly document this multi-hop tracing, which they term ``indirect exposure'' or ``indirect risk,'' following funds through intermediary addresses outward from known illicit entities: Chainalysis, \url{https://www.chainalysis.com/blog/cryptocurrency-risk-blockchain-analysis-indirect-exposure/}; TRM Labs, \url{https://www.trmlabs.com/resources/blog/key-considerations-for-evaluating-indirect-risk-on-the-blockchain}; and Crystal, \url{https://crystalintelligence.com/blockchain-analytics-tool/}.}}. Since these tools rely on flow-based analytics that track entire transaction chains rather than just the origin of an offense, the absence of a flag provides a strong baseline of \secreview{operationally} benign status\secreview{, that is, benign under the coverage of these providers and the filters described below rather than in an absolute sense}. \secreview{Because these providers propagate labels along transaction chains outward from confirmed illicit seeds, their reach extends beyond individually inspected addresses to the intermediaries and layering wallets connected to known illicit entities. This propagation is nonetheless bounded by each provider's detection thresholds and can miss intermediaries just past them, so the absence of a flag is a strong first screen rather than proof of complete coverage. We cannot state a precise coverage percentage, as the providers' methods and underlying intelligence are proprietary and no ground-truth set of truly-illicit-but-unflagged addresses exists against which coverage could be measured; we treat this residual as a limitation and add the safeguards below, the Proximity Exclusion Filter as a conservative buffer against provider false negatives and the behavioral legitimacy markers as a largely independent signal.} To ensure the integrity of this class and further mitigate potential label noise, we implemented a \textit{Proximity Exclusion Filter}. This filter systematically removes any candidate address within a 2- to 3-hop transaction radius (i.e., connected via up to three direct or indirect transaction edges) of known illicit clusters to eliminate undetected intermediaries. \secreview{The 2- to 3-hop bound is deliberate. Below two hops the filter would reach only the wallets in direct contact with an illicit node and leave the intermediaries two and three hops away, the layering positions through which funds are routed, inside the Normal pool as undetected accomplices. Beyond three hops the neighborhood on a network of nearly 55 million addresses expands rapidly, and a growing share of active wallets becomes reachable from some illicit node, so an unbounded radius would risk removing legitimate candidates and shrinking the Normal class toward emptiness. A 2- to 3-hop buffer therefore matches the depth of the layering paths the framework targets while preserving a usable benign pool.} Furthermore, an exploratory data analysis of the resulting Normal subset confirms behavioral legitimacy markers consistent with regulated retail activity. These markers include frequent interactions with KYC-compliant Centralized Exchanges (CEX) and a high incidence of long-term wallet status, both of which are systematically underrepresented in the illicit categories.
    
    \item \textit{\textbf{Cybercrime}}: addresses identified as direct perpetrators of malicious on-chain activities. These labels are sourced from forensic reports by blockchain security firms (e.g., SlowMist, PeckShield) and community alerts. Specifically, we collect them through automated ingestion of public threat-intelligence reports posted on Etherscan and on partner blockchain security firms' disclosures. This category captures the active originators of illicit funds, including:
    \begin{itemize}
        \item Exploits \& Hacks: addresses linked to DeFi protocol breaches or exchange thefts.
        \item Scams \& Phishing: addresses associated with fraudulent schemes or social engineering attacks.
\end{itemize}
     
 \item \textit{\textbf{Blocklisted}}: addresses subject to administrative interdiction or legal enforcement. Unlike the \textit{Cybercrime} category, which is defined by active behavior, this category is defined by regulatory status and encompasses both public and private enforcement mechanisms. We extract these labels by decoding the on-chain \texttt{addBlocklist} events of the USDT and USDC stablecoin contracts directly from blockchain logs, returning the same address set that the issuers' compliance teams act on.

    \begin{itemize}
        \item \textit{Sanctioned}: addresses designated by government bodies, specifically the OFAC Specially Designated Nationals (SDN) \citep{andersen2022sanctioned}.
        \item \textit{Frozen}: addresses blocked at the smart contract level by stablecoin issuers, rendering the assets non-transferable.
    \end{itemize}
\end{itemize}

In the resulting modeling dataset, \textit{Normal} wallets represent approximately 48.7\% of observations, while \textit{Cybercrime} and \textit{Blocklisted} accounts comprise 36.5\% and 14.8\%, respectively.

\subsection{Feature Engineering}
\label{subsec:feature_engineering}

Since money laundering on blockchains manifests through complex behavioral and relational patterns \citep{shojaeinasab2024decoding}, we construct a structured feature extraction process that quantifies wallet activity across multiple dimensions. 


After assembling the engineered attributes for each wallet, we organize them in a tabular format, with each row corresponding to an address and each column representing an engineered feature (details are provided in Table~\ref{tab:full_feature_list} of Appendix~\ref{App:features}). We group the resulting 68 features into four distinct categories:

\begin{itemize}
    \item \textbf{Interaction Features:} Describe direct wallet–protocol relationships (e.g., DeFi, Centralized Exchanges (CEX)).
    \item \textbf{Derived Network Features:} Track multi-hop fund propagation and indirect connections.
    \item \textbf{Transfer-based Features:} Characterize direct value flow and volume thresholds.
    \item \textbf{Temporal and Direct Features:} Capture timing patterns and intrinsic account properties.
\end{itemize}

\subsubsection{Interaction Features}
Feature construction begins by identifying the major on-chain services and protocols that wallets interact with, including swap protocols, lending platforms, and staking pools. Each represents a distinct functional layer of the Web3 ecosystem. By explicitly labeling these services and linking each transaction to the corresponding protocol, we construct features that capture how wallets route funds through intermediaries.

Table~\ref{tab:service_labels_summary} summarizes three key service categories. The ``Wallets'' column reports the number of addresses that activate each feature. Note that the sum of these counts exceeds the total number of unique wallets, as a single address may interact with multiple service types (e.g., using both Uniswap and Aave).

\begin{table}[H]
\centering{
\setlength{\extrarowheight}{2pt}
\renewcommand{\arraystretch}{1.25}
\begin{tabular}{p{3.18cm} p{6cm} p{2.5cm} r}
\hline
\textbf{Service Label} & \textbf{Description} & \textbf{Features} & \textbf{Wallets} \\
\hline
\textbf{Swap Protocols} & Permissionless token-exchange platforms such as Uniswap or Curve.  &  \texttt{\small sentToSwap}, \texttt{\small receivedFromSwap} & 44,312 \\[4pt]
\textbf{Lending Platforms} & Protocols enabling borrowing and collateralized lending, such as Aave or Compound. 
 & \texttt{\small sentToLending}, \texttt{\small receivedFromLending} & 28,904 \\[4pt]
\textbf{Staking Pools} & Smart contracts allowing users to lock tokens for yield or network participation.
 &  \texttt{\small sentToStake}, \texttt{\small receivedFromStake} & 12,587 \\
\hline
\end{tabular}
}
\caption{Service interaction labels. A wallet is counted in multiple categories if it interacts with multiple protocol types.}
\label{tab:service_labels_summary}
\end{table}

\subsubsection{Derived Network Features}
While interaction features capture direct relationships, derived network features capture how illicit funds propagate across the network after the initial transaction. These features track movements across multiple hops, including second- and third-degree connections to flagged wallets.

\secreview{In this context, we explicitly characterize these proximity features as known-bad-proximity screening signals. The inclusion of these features represents a deliberate emulation of the initial screening step routinely performed by operational compliance systems in real-world scenarios. It is important to note, as a scoping consideration, that the intelligence sources used to derive these signals share the same labeling universe as our ground-truth dataset.}

\begin{figure}[H]
    \centering 
    \includegraphics[width=1.05\linewidth]{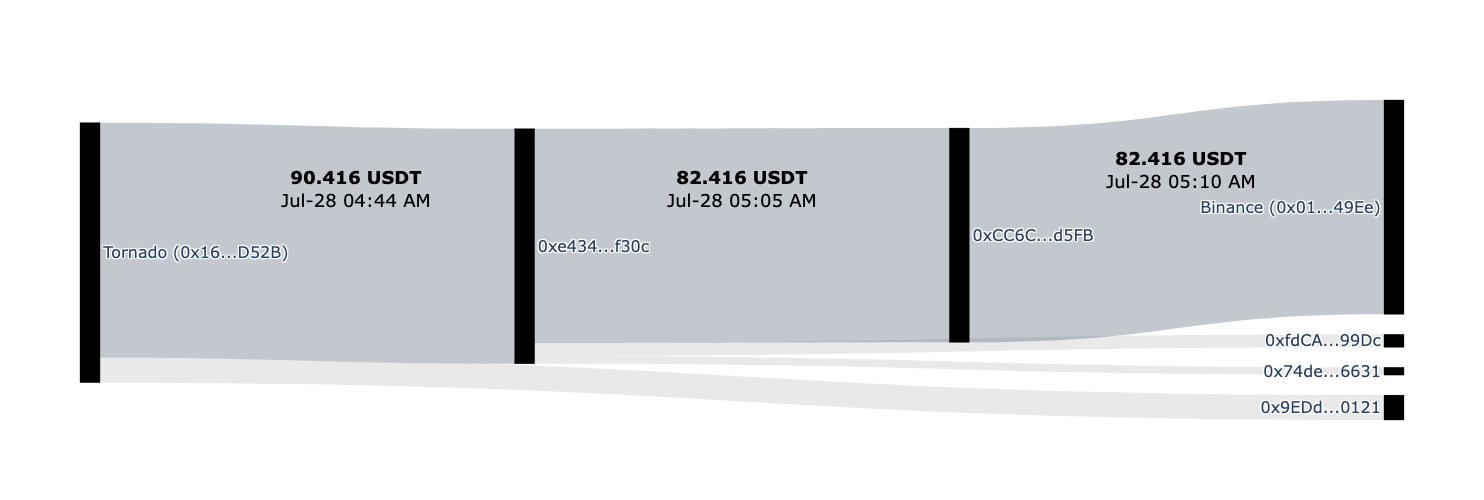}
    \caption{Sankey diagram of a multi-hop transfer. Funds are withdrawn from Tornado Cash, passed through intermediary wallets to obscure the trail, and rapidly deposited into a centralized exchange (Binance) within a 26-minute window. The prominent grey path highlights the primary volume flow, illustrating the ``layering'' behavior captured by second-degree network features.}
    \label{fig:sankey_flow}
\end{figure}

To illustrate multi-hop exposure, Figure~\ref{fig:sankey_flow} details a real-world laundering pattern observed in the dataset. The sequence begins at 04:44 UTC with a withdrawal of approximately 90 USDT from the Tornado Cash mixer (\texttt{\small 0x16...D52B}) to an intermediary wallet (\texttt{\small 0xe434...f30c}). In a typical obfuscation attempt, a portion of the funds is siphoned off or paid as fees, while the bulk (82.416 USDT) is forwarded twenty minutes later to a second intermediary wallet (\texttt{\small 0xCC6C...d5FB}). Finally, at 05:10 UTC, less than 30 minutes after the initial withdrawal, the funds are deposited into a Binance wallet (\texttt{\small 0x01...49e}) for off-ramping\footnote{Off-ramping refers to the process of converting digital assets back into fiat currency or tangible goods, serving as the exit point from the blockchain ecosystem.}. This rapid, multi-hop transfer\footnote{Full transaction hashes: \texttt{0x5f35d27d5d17f1249f6245544f0426aeb091fdaafb16db956afa636c4ca0111d}, representing the withdrawal from the Tornado Cash router to the first intermediate wallet; \texttt{0x41c2e200e8d078883f88aceaf117acb9230a25e7ddce46378f4128e50b74abf1}, recording the transfer from the first intermediate wallet to a downstream recipient approximately twenty minutes after the mixer withdrawal; and \texttt{0x4a5be4a5df2099cc53eedaba8b9185c465d0d8dc55591475d971cf1ea46fccaa}, representing the subsequent transfer from the downstream wallet to the final destination address.} during off-peak hours exemplifies the structural ``hop'' patterns our derived features are engineered to detect.

\begin{figure}[H]
    \centering
    \begin{subfigure}[b]{0.48\textwidth}
        \includegraphics[width=\textwidth]{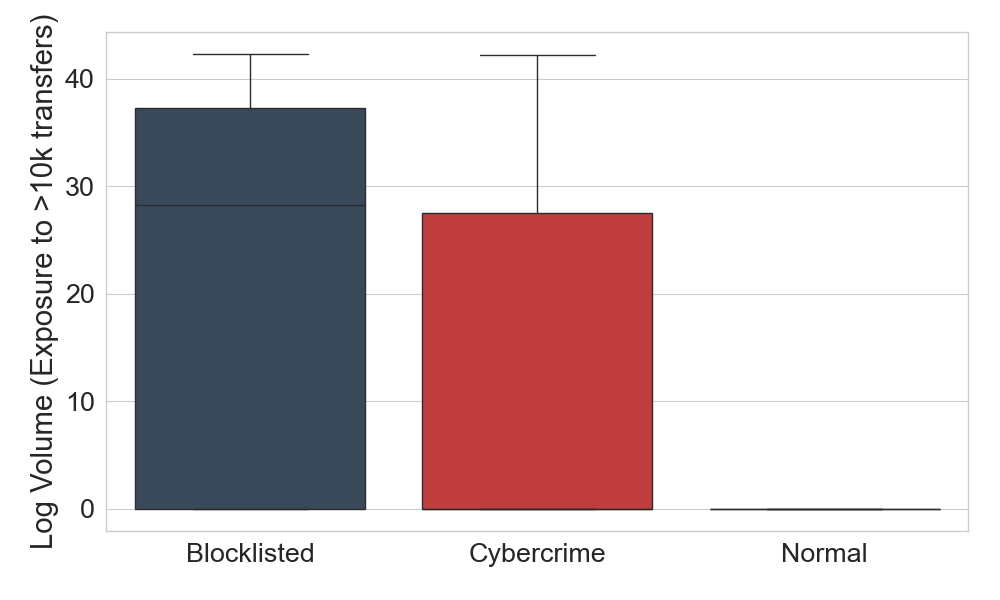}
        \caption{Indirect High-Value Exposure (Log Scale)}
    \end{subfigure}
    \hfill
    \begin{subfigure}[b]{0.48\textwidth}
        \includegraphics[width=\textwidth]{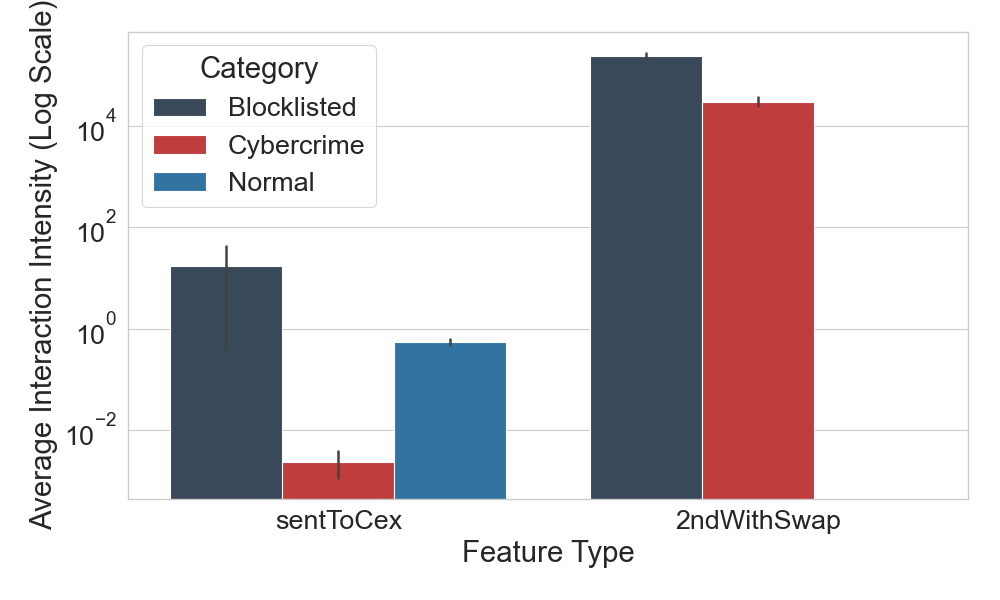}
        \caption{Protocol Interaction Profile}
    \end{subfigure}
    \caption{Exploratory plots. (a) Distribution of \texttt{2ndWithOver10k} shows that illicit wallets are structurally embedded in high-volume networks, unlike normal users. (b) Inverse preference for CEX vs. DeFi Swaps.}
    \label{fig:eda_plots}
\end{figure}

To appreciate the role of interaction and derived network features, Figure~\ref{fig:eda_plots} visualizes the distribution of key indicators. The left panel demonstrates the structural difference in indirect risk exposure: while legitimate users have zero exposure to high-value laundering flows, Blocklisted and Cybercrime wallets show a distinct distribution of second-degree connections to transfers exceeding \$10k (\texttt{2ndWithOver10k}). The right panel highlights the operational divergence: legitimate users favor Centralized Exchanges (\texttt{sentToCex}), whereas illicit actors predominantly route funds through decentralized swap protocols.

\subsubsection{Transfer-based Features}
Transfer-based features capture the direct flow of value \textit{into} and \textit{out} of a wallet. These include threshold-based indicators, such as transfers exceeding \$1,000, \$5,000, and \$10,000. Their design draws on the classic AML literature, which notes that illicit funds typically move through identifiable entry and exit channels during the placement and integration stages of laundering. Rather than assuming these stages \textit{a priori}, the features capture transactional patterns consistent with them, allowing the model to learn whether such movements signal elevated risk.

\subsubsection{Temporal and Direct Features}
Temporal behavioral features track time-based patterns such as daily transfer bursts, repeated identical-value movements, and rapid reciprocals. These highlight abnormal rhythms often aligning with laundering practices, such as sudden high-frequency inflows. To illustrate, the address \texttt{\small 0x03d09ec664f9241b23223240a06079300de1b14d} sends a sequence of repeated 50,000~USDT transfers to the recipient \texttt{\small 0xeFd2fd5c18093030E15a08fF8799BEC9c612Ec4f}. On 2025-08-08, the recipient receives four such transfers in less than fifteen minutes, totaling 200,000~USDT. Table~\ref{table:temporal_csv_example} provides a concrete example of the temporal anomalies these features are designed to capture.

\begin{table}[H]
\centering
\begin{tabular}{p{10.3cm}  p{2cm} r}
\hline
\textbf{Transaction} &  \textbf{Timestamp} &  \textbf{USDT}  \\
\hline
\small{0xeb073c65c481114289e8aa63d86ca36ae79854eb270b58ce12961c4d41666124}  & \small{2025-08-08 03:04:47 UTC} & 50{,}000  \\
\small{0x16a3e2bb4c73bc50324b3ba24ae391ae3ead13e90deed2ea835f3c1fb4e780cb}  & \small{2025-08-08 03:04:59 UTC} & 50{,}000  \\
\small{0x74b8321b5c8817e0dc00c21907e8a85673f1762caa21a3d52328ca6421db6e59}  & \small{2025-08-08 03:20:11 UTC} & 50{,}000 \\
\small{0xb4a930b2e5aaf8d09055fea6eda1a134cb1524ba6be86418119eb1714d4329a1}  & \small{2025-08-08 03:20:23 UTC} & 50{,}000  \\
\hline
\end{tabular}
\caption{Temporal excerpt showing repeated identical-value inflows of 50,000~USDT from \texttt{\small 0x03...} to \texttt{\small 0xeF...}.}

\label{table:temporal_csv_example}
\end{table}

Finally, direct features capture intrinsic properties of a wallet that do not depend on interactions with other addresses. These include whether the address is an externally owned account (EOA)\footnote{An EOA refers to a standard user-controlled wallet managed by a private key, in contrast to a smart contract account whose behavior is defined by on-chain code.}, whether its contract code is verified\footnote{A verified smart contract publishes its source code and metadata on block explorers such as Etherscan, enabling public inspection and confirmation that the deployed bytecode matches the published source.}, and indicators of prolonged activity\footnote{Prolonged activity refers to wallets that remain active over extended periods, as opposed to short-lived or disposable addresses often used in obfuscation or layering schemes.}. These properties help distinguish ordinary user wallets from ephemeral or synthetic addresses frequently observed in laundering operations.

\subsection{Graph Construction}
While the feature-based classifiers described previously excel at identifying individual suspicious behaviors, they treat each wallet as an independent entity, potentially overlooking the complex, collaborative nature of money laundering networks. To bridge this gap, we complement the tabular analysis with a graph approach, explicitly modeling the transaction ecosystem to capture how illicit flows propagate between accounts.

To leverage the relational structure of the ecosystem, we model the data as a weighted directed graph $\mathcal{G} = (\mathcal{V}, \mathcal{E})$.
\begin{itemize}
    \item \textbf{Nodes ($\mathcal{V}$):} The set of unique wallet addresses identified in the \textit{StableAML} dataset.
    \item \textbf{Edges ($\mathcal{E}$):} For each ordered pair of wallets $(A,B)$ with at least one observed transfer, we define a directed edge $e_{A \rightarrow B}$. The edge weight $w_{A \rightarrow B}$ equals the total stablecoin volume transferred from $A$ to $B$, aggregated over the full temporal span of the dataset.
    
\end{itemize}

We formulate the detection task as supervised multiclass node classification, mapping each node $v$ to a risk category $y \in \{\text{\textit{Normal}}, \text{\textit{Cybercrime}}, \text{\textit{Blocklisted}}\}$. To integrate behavioral insights, we initialize each node $v \in \mathcal{V}$ with a feature vector $\mathbf{x}_v \in \mathbb{R}^{68}$, corresponding to the engineered attributes defined in Section~\ref{subsec:feature_engineering}.
Consistent with the tabular setting, we construct the graph over the curated set of labeled wallets and perform a wallet-level train–test split on this fully labeled node set, as explained in the following section. The splits are disjoint, ensuring that no wallet appears in both training and test sets. As both the graph structure and node features are cumulative aggregates over the full observation window, the task is formulated as static multiclass node classification.

This design choice is deliberate and reflects the operational constraint we aim to simulate. In the broader Web3 ecosystem, native transaction graphs are increasingly fragmented by privacy-preserving protocols, mixers, and cross-chain bridges, leaving stablecoin ledgers as the most persistent and auditable layer accessible to compliance authorities. By restricting the graph to stablecoin edges over the labeled wallet set, we evaluate the message-passing models within this realistic \textit{Compliance Tunnel} rather than under an idealized fully observed transaction network. The resulting sparsity, with edge density below 0.01, is therefore not an artifact of dataset curation but a structural property of the regulatory environment in which AML detection on stablecoins is expected to operate. Consequently, the comparison between graph-based and feature-based methods reported in Section~\ref{Sec:results} should be interpreted as evidence of relative resilience to topological fragmentation, rather than as a general statement on the superiority of one architectural family over another.

\section{Modeling Methodology}\label{Sec:models}
In this section, we formalize the multiclass wallet classification task and outline the evaluation framework. The task is a three-class classification problem ($K=3$), labeling wallets that transfer USDT or USDC according to the categories defined in Section~\ref{Sec:dataset}. Additionally, we provide results for a binary classification setup ($K=2$) to further analyze the models' efficacy in distinguishing between broader risk categories.

Given the strong class imbalance, with \textit{Cybercrime} and \textit{Blocklisted} wallets comprising a small fraction of the raw ecosystem, we evaluate all models using the Macro-F1 Score (F1\textsubscript{Macro}). This metric treats each class equally and emphasizes the balance between precision and recall, which is more informative than overall accuracy under label imbalance. While we also report the \textit{Area Under the Receiver Operating Characteristic} (AUROC) for completeness, it is considered a secondary metric here, as AUROC tends to yield overly optimistic results in the presence of highly skewed class distributions. In AML settings, this balance is critical: false negatives carry substantially higher regulatory and operational costs than false positives. Missing an illicit wallet can enable the continuation of criminal flows, expose institutions to sanctions, and violate compliance obligations, as highlighted by the Financial Action Task Force risk-based framework \citep{force2021updated}. Prior AML research similarly argues that recall deserves particular attention when the goal is to identify rare but high-risk entities \citep{shojaeinasab2024decoding}. 

Within each model family we select architectures that are widely adopted in tabular and graph learning, ensuring that any performance gap reflects family-level inductive biases rather than implementation idiosyncrasies. For the linear family, we adopt Logistic Regression with $\ell_1$/$\ell_2$ penalty selection via grid search, the standard supervised baseline for high-dimensional tabular classification. For the tree-ensemble family, we evaluate four canonical implementations, Random Forest, XGBoost, CatBoost, and LightGBM, which collectively span the bagging and gradient-boosting designs employed in industry-grade fraud-detection systems. For deep neural networks, we adopt a fully connected architecture with ReLU activations and dropout, a representative configuration for tabular deep learning. For graph neural networks, we evaluate both static (GraphSAGE) and dynamic (EvolveGCN, in both -H and -O variants) architectures, which together span the inductive-static and temporal-dynamic axes of the GNN literature most relevant to financial transaction networks. This family-wise selection is intentionally representative rather than exhaustive; comprehensive ablation across every published variant is beyond the scope of the present study and is discussed as future work in Section~\ref{Sec:conclusion}.

We split the dataset into training and test sets using an 80/20 partition, resulting in 13,146 training samples and 3,287 independent test samples. The test portion serves as a fully isolated holdout set for the final assessment of generalization performance.

The resulting class distribution within these partitions allows the model to learn detailed risk patterns that would otherwise be statistically overwhelmed by the majority class, without relying on synthetic data generation. After selecting the best configuration, we retrain the model on the full training split and evaluate it once on the holdout test set. The complete set of optimized hyperparameters for all models appears in Appendix~\ref{sec:hyperparams}.

\subsection{Logistic Regression}

We use logistic regression (LR) as an interpretable baseline to assess the discriminatory power of the engineered features under linear decision boundaries. We adopt the multinomial logistic regression formulation, which generalizes logistic regression to multiclass problems by modeling the joint probability distribution across all $K$ classes simultaneously.

The predicted probability that a wallet with feature vector $\mathbf{x} \in \mathbb{R}^{68}$ belongs to class $k$ is computed using the softmax function:

\begin{equation*}
P_k(\mathbf{x}) = \frac{\exp(\beta_{k}^{\top}\mathbf{x} + \beta_{0,k})}{\sum_{j=1}^{K} \exp(\beta_{j}^{\top}\mathbf{x} + \beta_{0,j})}, \quad k \in \{0, 1, 2\}
\end{equation*}

where $\beta_{k} \in \mathbb{R}^{68}$ denotes the class-specific coefficient vector, $\beta_{0,k}$ represents the intercept for class $k$, and $K=3$ corresponds to the total number of risk categories. The final class assignment is determined by selecting the category with the highest probability mass:

\begin{equation*}
\hat{y}(\mathbf{x}) = \operatorname*{argmax}_{k} P_k(\mathbf{x})
\end{equation*}

Model parameters are estimated by minimizing the categorical cross-entropy loss. We utilize the SAGA solver to optimize the objective function, as it provides efficient convergence for datasets with high-dimensional feature spaces and supports the $L_1$ regularization penalties explored in our grid search.

\subsection{Tree Ensembles}
Ensemble learning aggregates predictions from multiple base estimators to improve generalization and robustness compared to a single model. We leverage this paradigm to capture the complex, non-linear feature interactions characteristic of money laundering patterns. Specifically, we benchmark four tree ensembles selected for their proven efficacy in handling heterogeneous behavioral signals and sparse transaction data \citep{chen2018machine,shojaeinasab2024decoding}: Random Forest (RF) and three Gradient Boosting Machines (GBMs): XGBoost, LightGBM, and CatBoost.

RF employs bagging to reduce variance by aggregating predictions from an ensemble of independent decision trees. The predicted probability that a wallet with features $\mathbf{x}$ belongs to class $k$ is computed as the average of the empirical class proportions produced by each of the $M$ individual trees, where $P_m(y=k \mid \mathbf{x})$ denotes the fraction of training samples of class $k$ in the terminal leaf of tree $m$. To optimize the decision nodes, we employ the Gini impurity index as the splitting criterion, and the final class assignment follows:

\begin{equation*}
    \hat{y}(\mathbf{x}) = \operatorname*{argmax}_{k} P_k(\mathbf{x}).
    \label{eq:rf_combined}
\end{equation*}

In contrast to the independent training of RF, GBMs construct an additive model in a forward stage-wise fashion. They iteratively fit new decision trees to the negative gradient of the loss function associated with the current ensemble's prediction, effectively performing gradient descent in function space to minimize a global objective.

To address the multiclass nature of our task, we adopt the native multiclass formulation implemented in XGBoost, LightGBM, and CatBoost. In this setting, the ensemble produces a vector of class-specific scores (logits) $F_k(\mathbf{x})$, obtained by aggregating the contributions of $M$ trees:

\begin{equation*}
    F_k(\mathbf{x}) = \sum_{m=1}^{M} f_{m,k}(\mathbf{x}).
    \label{eq:gbm_sum}
\end{equation*}

The class probabilities are computed using the softmax function:

\begin{equation*}
    P_k(\mathbf{x}) = \frac{\exp(F_k(\mathbf{x}))}{\sum_{j=1}^{K} \exp(F_j(\mathbf{x}))},
    \label{eq:gbm_softmax}
\end{equation*}

and the predicted label is given by

\begin{equation*}
    \hat{y}(\mathbf{x}) = \operatorname*{argmax}_{k} P_k(\mathbf{x}).
    \label{eq:gbm_pred}
\end{equation*}

Model parameters are optimized by minimizing the multinomial cross-entropy loss:

\begin{equation*}
    \mathcal{L} = -\frac{1}{n} \sum_{i=1}^{n} \sum_{k=1}^{K}
    y_{i,k} \log P_k(\mathbf{x}_i).
    \label{eq:gbm_loss}
\end{equation*}

The complete configuration of grid search spaces and optimal hyperparameters is detailed in Appendix~\ref{sec:hyperparams}.

\subsection{Deep Neural Network} 
To capture high-order non-linear feature interactions, we implement a Deep Neural Network (DNN) architecture consisting of two fully connected hidden layers. Non-linearity is induced via Rectified Linear Unit (ReLU) activations, defined as $f(x) = \max(0, x)$. To improve generalization and prevent overfitting on the high-dimensional feature space, we incorporate dropout regularization, which randomly masks a fraction of neuronal connections during the training phase.

The final hidden representation is linearly projected to a vector of unnormalized logits $\mathbf{z} \in \mathbb{R}^K$ and normalized using the softmax function:
\begin{equation}
    P_k(\mathbf{x}) = \frac{\exp(z_k(\mathbf{x}))}{\sum_{j=1}^{K} \exp(z_j(\mathbf{x}))}.
\end{equation}

The predicted class label $\hat{y}$ is determined by selecting the category with the highest probability mass, as follows:

\begin{equation*}
    \hat{y}(\mathbf{x}) = \operatorname*{argmax}_{k} P_k(\mathbf{x}).
    \label{eq:softmax_pred}
\end{equation*}

Model parameters are optimized using the Adam algorithm by minimizing the categorical cross-entropy loss over the $n$ training samples:

\begin{equation*}
    \mathcal{L} = - \frac{1}{n} 
    \sum_{i=1}^n \sum_{k=1}^K 
    y_{i,k} \log P_k(\mathbf{x}_i),
    \label{eq:crossentropy_total}
\end{equation*}

where $y_{i,k} \in \{0,1\}$ is the binary indicator for the true class membership of sample $i$, and $P_k(\mathbf{x}_i)$ denotes the predicted class probability.

\subsection{Graph Neural Network Architecture}

\label{subsec:gnn_model}
To evaluate the efficacy of graph-based approaches within fragmented stablecoin networks, we employ both static and dynamic message-passing architectures.

We adopt the GraphSAGE architecture \cite{hamilton2017inductive}, which learns neighborhood aggregation functions rather than node-specific embeddings. The inductive design allows the model to generalize by aggregating local neighborhood features, making it well-suited for large and evolving transaction networks. \secreview{Consistent with the Compliance Tunnel setting of Section~3.3, our primary results confine GNN message passing to the labeled wallet subgraph. As a robustness check, we additionally evaluate an extended variant that routes messages through the first- and second-degree neighborhood of the labeled wallets; because the architecture uses two message-passing layers, this expansion supplies the full two-hop receptive field a two-layer model can exploit. The extended-scope results are detailed by class in Table~\ref{table_three_class_results} of Appendix~C and summarized in Section~5.2. Expanding the scope narrows the gap against the tree ensembles for the \textit{Normal} and \textit{Cybercrime} classes but leaves the \textit{Blocklisted} metrics stagnant, consistent with the restricted, linear paths of frozen or sanctioned wallets.}

The architecture consists of two spatial convolution layers equipped with neighbor sampling to handle graph sparsity. At layer $l$, the model updates the representation of node $v$ by concatenating its current embedding with an aggregated summary of its local neighborhood. We utilize the mean aggregator for its efficiency and permutation invariance. The update rule is formally defined as:

\begin{equation*}
    h_v^{(l)} = \sigma \left( W^{(l)} \cdot \left[ h_v^{(l-1)} \, \Bigg\| \, \frac{1}{|\mathcal{N}(v)|} \sum_{u \in \mathcal{N}(v)} h_u^{(l-1)} \right] \right),
    \label{eq:graphsage_update}
\end{equation*}

where $h_v^{(0)} = \mathbf{x}_v$ corresponds to the initial feature vector described in Section~\ref{subsec:feature_engineering}, $\|$ denotes the concatenation operation, and $W^{(l)}$ is a learnable weight matrix. The term within the summation computes the element-wise mean of the neighbor embeddings, capturing the average behavioral profile of a wallet's direct transactors. $\sigma$ represents the non-linear activation function (typically a Rectified Linear Unit (ReLU)).

After $L=2$ layers of aggregation, the final node embeddings $h_v^{(L)}$ are projected to the output space corresponding to the three target classes via a linear transformation. The class probabilities are obtained by applying the softmax function:
\[
P_k(h_v^{(L)}) = \frac{\exp(z_{v,k})}{\sum_{j=1}^{K} \exp(z_{v,j})},
\]
The predicted label is then given by $\hat{y}_v = \operatorname*{argmax}_{k} P_k(h_v^{(L)})$.

To ensure our evaluation is not constrained by the static nature of GraphSAGE, we additionally incorporate EvolveGCN \cite{pareja2020evolvegcn} to capture temporal network dynamics. At any given time step $t$, the node representations for layer $l$ are updated using a standard graph convolution:
$$H_t^{(l)} = \sigma \left( \widehat{A}_t H_t^{(l-1)} W_t^{(l)} \right)$$
where $\widehat{A}_t$ is the normalized adjacency matrix at time $t$, $H_t^{(l-1)}$ represents the node features from the previous layer, and $W_t^{(l)}$ is the weight matrix.

The defining characteristic of EvolveGCN is the temporal evolution of this weight matrix $W_t^{(l)}$, which is updated sequentially using a recurrent neural network. Within this dynamic family, we evaluate both the -H and -O variants to systematically assess the impact of feature-based noise. In the EvolveGCN-H variant, the weights are updated using a Gated Recurrent Unit (GRU) that takes the current node features $H_t^{(l-1)}$ as input to guide the evolution: 

$$W_t^{(l)} = \text{GRU} \left( H_t^{(l-1)}, W_{t-1}^{(l)} \right)$$

Conversely, the EvolveGCN-O variant utilizes a Long Short-Term Memory (LSTM) network to evolve the weights based solely on historical weights, completely bypassing the node feature representations during the temporal update:
$$W_t^{(l)} = \text{LSTM} \left( W_{t-1}^{(l)} \right)$$ 

Theoretically, the -O variant's independence from node hidden representations during weight updates may offer greater resilience to the extreme topological fragmentation present in our dataset, which can otherwise introduce significant noise during feature-based message passing. Together, these static and dynamic architectures allow us to empirically investigate whether the performance bottleneck in stablecoin AML stems from architectural limitations or from a fundamental reliance on persistent edge connectivity in deliberately obfuscated environments.

\section{Results and Discussion} \label{Sec:results}

This section presents the empirical findings of the proposed framework and discusses their implications for detecting suspicious activity in stablecoin ecosystems. We first evaluate model performance in the full three-class setting. We then examine feature importance to identify which behavioral and relational attributes most strongly influence predictive performance. Building on these results, the following subsection analyzes how these features align with known suspicious activity typologies, revealing structural differences between illicit and restricted wallet behaviors. We further interpret the detected patterns, linking them to characteristic money-laundering mechanisms and transaction-flow signatures. Finally, for comparability with prior work, we evaluate a simplified binary classification setup that collapses all suspicious wallets into a single class, providing an additional perspective on the expressiveness and robustness of the engineered feature space.

To facilitate reproducibility, the StableAML Dataset used in this study is available at \url{https://github.com/finos-labs/dtcch-2025-OpenAML} (OpenAML Project under the Linux Foundation FINOS initiative).

\subsection{Per-Class Performance Analysis}

We begin by examining the detailed breakdown of classification performance across the three target classes: \textit{Normal}, \textit{Cybercrime}, and \textit{Blocklisted}. Table~\ref{tab:per_class_results} presents the precision, recall, F1-score, and OvR AUROC for each category.

A critical observation concerns the divergence between AUROC and F1-score, particularly for the linear baseline. While LR achieves AUROC values above $0.93$ for all classes, its recall for the \textit{Blocklisted} class drops to $0.49$. This confirms that AUROC can be overly optimistic in imbalanced settings, masking significant false negatives. Consequently, we prioritize the F1-score and Recall as the primary indicators of operational utility.

The results reveal clear distinctions in learnability across risk typologies:

\begin{itemize}
    \item \textbf{Normal \& Cybercrime:} Tree ensembles achieve near-optimal performance for these categories, with F1-score exceeding $0.98$. This indicates that the engineered behavioral features effectively capture the consistent and often automated transaction patterns associated with both legitimate high-frequency users and professionalized cybercrime operations.
    
    \item \textbf{Blocklisted:} Classification of this category proves substantially more challenging. Although ensemble methods significantly outperform the LR baseline ($F1 \approx 0.95$ vs. $0.63$), the gap between precision and recall is most pronounced here. This difficulty reflects the heterogeneous nature of the Blocklisted class, which includes wallets frozen for diverse reasons, from sanctions evasion to theft recovery, resulting in less uniform transactional footprints than the active Cybercrime wallets.
\end{itemize}

Deep learning approaches (DNN and GNN) exhibit marked instability in this minority class. The GNN, in particular, struggles to recall Blocklisted entities ($Recall \approx 0.74$), lagging behind CatBoost ($Recall \approx 0.95$) by a wide margin. This suggests that in the absence of a dense transaction graph, deep learning models fail to recover the subtle signals that tree ensembles successfully extract from the tabular features.

\begin{table}[H]
\centering
\small
\setlength{\extrarowheight}{2pt}
\begin{tabular}{l l c c c c}
\hline
\textbf{Model} & \textbf{Class} & \textbf{AUROC} & \textbf{Precision} & \textbf{Recall} & \textbf{F1} \\
\hline
\multirow{3}{*}{Logistic Regression} 
 & Normal & 0.9453 & 0.7877 & 0.9988 & 0.8807 \\
 & Cybercrime & 0.9351 & 0.9375 & 0.7614 & 0.8403 \\
 & Blocklisted & 0.9353 & 0.8773 & 0.4896 & 0.6285 \\
\hline
\multirow{3}{*}{CatBoost (Best)} 
 & Normal & 0.9990 & 0.9927 & 1.0000 & 0.9963 \\
 & Cybercrime & 0.9991 & 0.9873 & 0.9831 & 0.9852 \\
 & Blocklisted & 0.9943 & 0.9579 & 0.9440 & 0.9509 \\
\hline
\multirow{3}{*}{EvolveGCN-O} 
 & Normal & 0.9922 & 0.9075 & 0.9968 & 0.9501  \\
 & Cybercrime & 0.9865 & 0.9207 & 0.9419 & 0.9312 \\
 & Blocklisted & 0.9602 & 0.8912 & 0.5782 & 0.7014 \\
\hline
\end{tabular}
\caption{Detailed per-class performance metrics on the test set. Note the significant drop in Recall for the Blocklisted class in non-ensemble models, despite high AUROC values.}
\label{tab:per_class_results}
\end{table}

The complete performance breakdown, demonstrating consistent high efficacy across different models, is detailed in Table~\ref{table_three_class_results} of Appendix~C.

\subsection{Overall Model Comparison}
\label{subsec:overall_results}

Aggregating performance across all categories, Table~\ref{tab:overall_results} summarizes the macro-averaged metrics for the evaluated models. The results demonstrate a clear hierarchy of model efficacy, with tree ensembles substantially outperforming both linear baselines and deep learning approaches.

\begin{table}[H]
\centering
\small
\begin{tabular}{l c c c c}
\hline
\textbf{Model} & \textbf{AUROC} & \textbf{Accuracy} & \textbf{F1} & \textbf{Recall} \\
\hline
Logistic Regression & 0.9385 & 0.8387 & 0.7831 & 0.7499 \\
Random Forest & 0.9976 & 0.9824 & 0.9714 & 0.9708 \\
LightGBM & 0.9976 & 0.9845 & 0.9755 & 0.9742 \\
\textbf{CatBoost} & \textbf{0.9974} & \textbf{0.9857} & \textbf{0.9775} & \textbf{0.9757} \\
XGBoost & 0.9974 & 0.9851 & 0.9766 & 0.9760 \\
DNN & 0.9617 & 0.9192 & 0.8708 & 0.8504 \\
GraphSAGE & 0.9683 & 0.8290 & 0.8048 & 0.7699 \\
EvolveGCN-O & 0.9796 & 0.9107 & 0.8608 & 0.8389 \\
\hline
\end{tabular}
\caption{Overall multiclass classification performance (macro-averaged metrics). CatBoost demonstrates the best balance between precision and recall across all risk categories.}
\label{tab:overall_results}
\end{table}

GBMs consistently achieve strong discrimination across all evaluation metrics, with average AUROC values exceeding $0.997$ and Macro-F1 scores ranging between $0.975$ and $0.979$. Performance is robust and well-balanced: the \textit{Normal} class exhibits near-perfect precision and recall, the \textit{Cybercrime} class is identified with F1 scores of approximately $0.98$, and even the heterogeneous \textit{Blocklisted} class achieves F1 scores of approximately $0.95$.

These results highlight the adaptability of tree ensembles to the specific properties of blockchain transaction data. As noted in broader machine learning literature, they frequently outperform deep neural networks on tabular datasets dominated by binary or dummy-encoded indicators, as they naturally partition the feature space along discrete signals without the need for extensive smoothing or manifold learning \citep{grinsztajn2022why}. Our engineered feature set contains many such behavioral flags (e.g., \texttt{isVerifiedContract}, \texttt{hasProxyBehaviour}), which these ensembles exploit effectively to isolate high-risk typologies.

In contrast, the deep learning approaches perform less competitively. While both DNN and GNN models achieve AUROC values above $0.98$, their Macro-F1 scores plateau around $0.80$--$0.87$. A central finding is the underperformance of GraphSAGE and EvolveGCN, which fail to improve upon the DNN baseline ($F1 \approx 0.86$ vs $0.87$).

\secreview{Furthermore, to investigate whether this limitation was merely an artifact of static modeling, we evaluated the dynamic EvolveGCN architecture. As hypothesized in Section~\ref{subsec:gnn_model}, the EvolveGCN-O variant, which evolves network weights independently of the noisy node features, outperformed both GraphSAGE and its feature-dependent -H counterpart (Macro-F1 $0.8608$ vs.\ $0.8048$ and $0.8349$, respectively; see Table~\ref{tab:overall_results} and Table~\ref{table_three_class_results}). However, despite this temporal adaptation, EvolveGCN-O still peaks at an F1-score of $0.9312$ for the Cybercrime class, lagging significantly behind the CatBoost model ($0.9852$). As a robustness check, we additionally re-ran the GNN suite with message passing extended to the first- and second-degree neighborhood beyond the labeled set. Under this expanded scope, GraphSAGE attains the highest Macro-F1 among the three architectures ($0.8670$), ahead of EvolveGCN-H ($0.8629$) and EvolveGCN-O ($0.8548$). Between the two EvolveGCN variants, EvolveGCN-H narrowly surpasses EvolveGCN-O, suggesting that once sufficient topological context is restored, conditioning weight updates directly on node features provides a marginal edge over feature-independent evolution. Even so, the extended EvolveGCN-H tops out at an F1-score of $0.9265$ for the Cybercrime class, still trailing CatBoost by a wide margin.}

This outcome reflects data constraints specific to the stablecoin ecosystem. The resulting transaction graph exhibits extreme sparsity (density $< 0.01$, where density denotes the fraction of observed edges relative to the maximum number of possible directed edges $|\mathcal{V}|(|\mathcal{V}|-1)$, excluding self-loops), primarily because the analysis is intentionally restricted to USDT and USDC transfers. Money laundering workflows frequently involve rapid swaps of stablecoins into volatile assets (e.g., ETH, WBTC) via decentralized exchanges to break the transaction chain. Since these intermediate cross-asset interactions are excluded from a stablecoin-only graph, the connectivity required for effective message passing is broken. Consequently, the GNN cannot capture the multi-hop layering patterns that are otherwise visible to the ensemble models via engineered aggregate features (e.g., \texttt{2ndWithSwap}).

\begin{figure}[H]
    \centering
    \begin{subfigure}[b]{0.49\textwidth}
        \includegraphics[width=\linewidth]{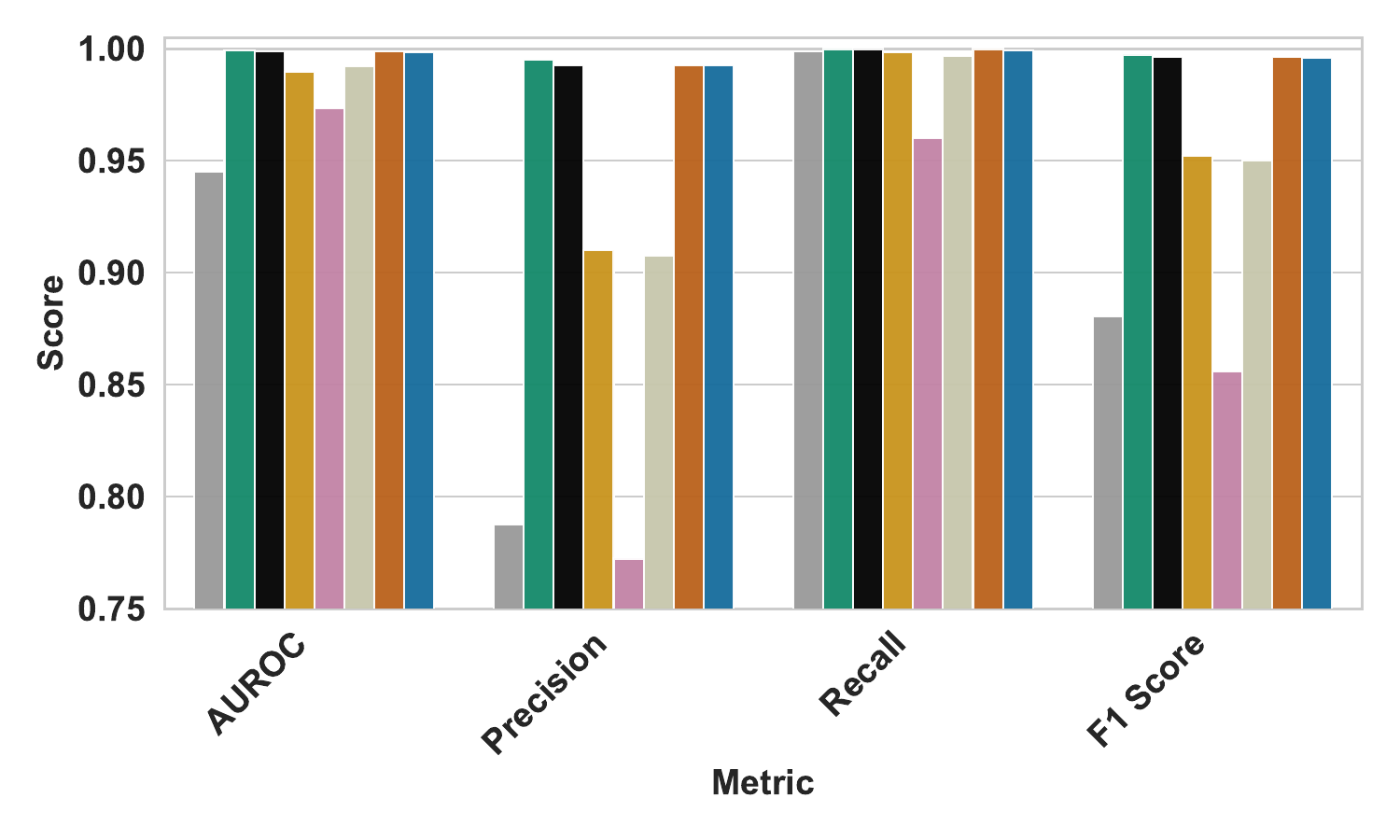}
        \caption{Normal}
    \end{subfigure}
    \hfill
    \begin{subfigure}[b]{0.49\textwidth}
        \includegraphics[width=\linewidth]{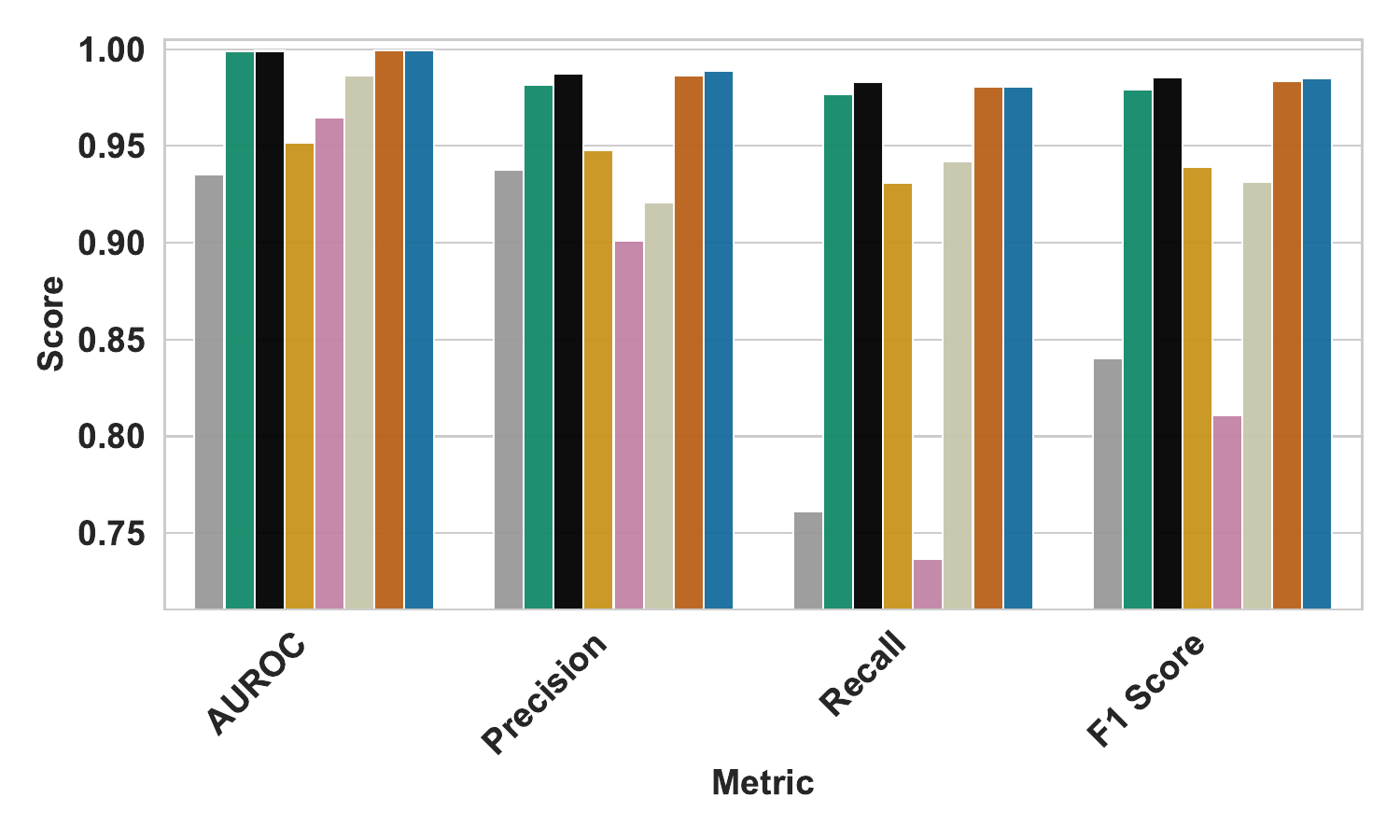}
        \caption{Cybercrime}
    \end{subfigure}

    \vspace{1em}

    \begin{subfigure}[c]{0.49\textwidth}
        \includegraphics[width=\linewidth]{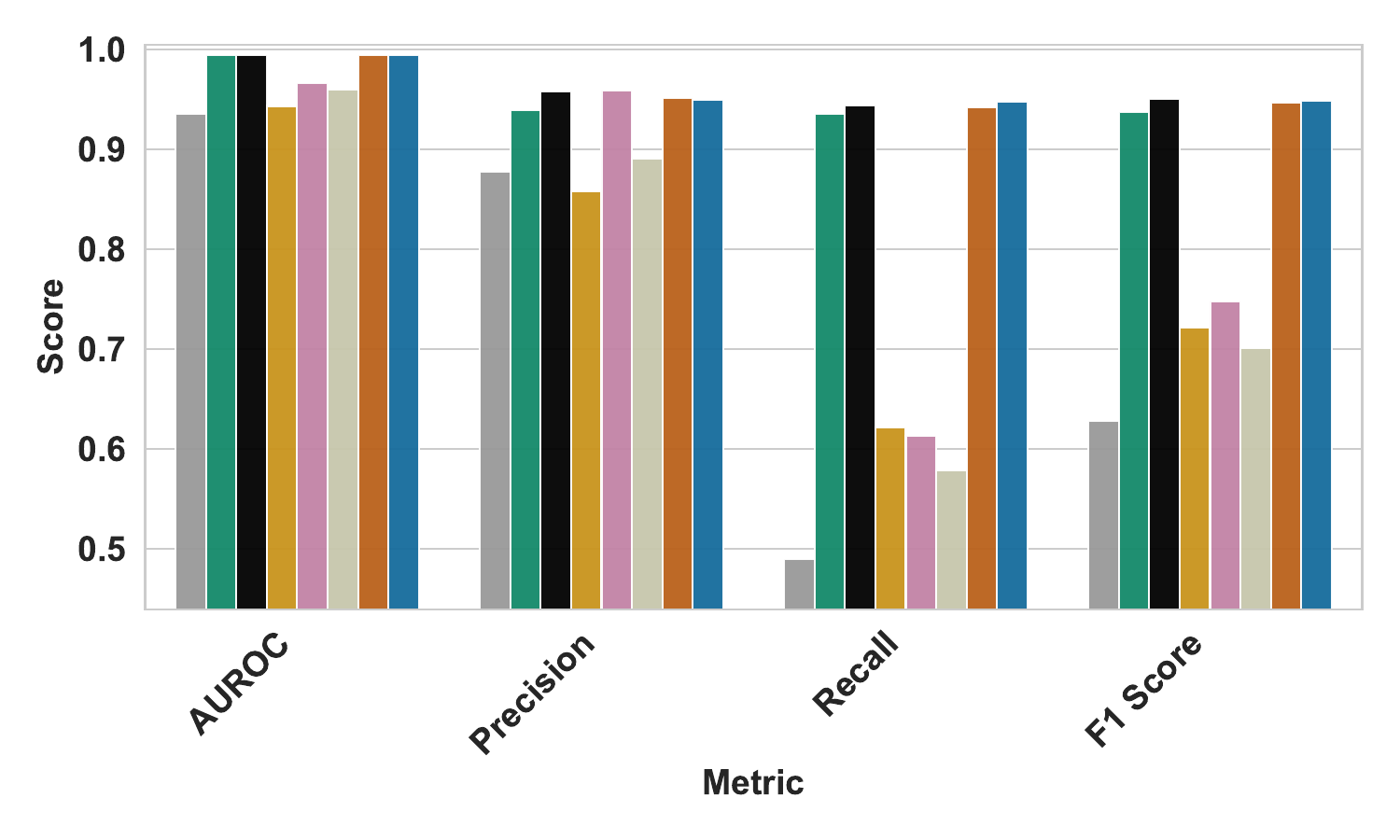}
        \caption{Blocklisted}
    \end{subfigure}
    \hfill
    \begin{subfigure}[c]{0.48\textwidth}
        \centering
        \includegraphics[width=0.48\linewidth]{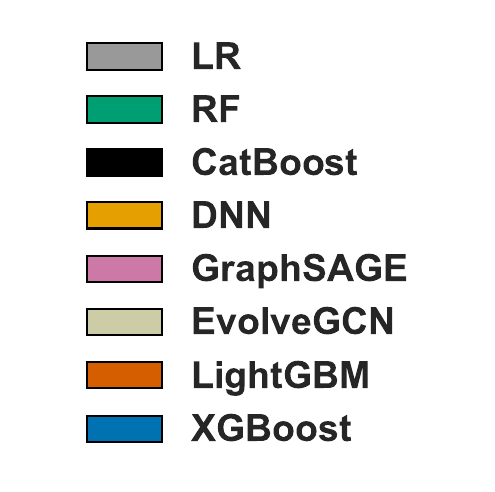}
    \end{subfigure}

    \caption{Per-class OvR performance curves for multiclass wallet classification, shown separately for the \textit{Normal}, \textit{Cybercrime}, and \textit{Blocklisted} classes.}

    \label{Fig:chart_class_level_multi}
\end{figure}

The detailed per-class performance analysis, summarized in Figure~\ref{Fig:chart_class_level_multi}, reveals clear architectural differences in how models handle distinct risk typologies. For the \textit{Normal} and \textit{Cybercrime} classes (Panels (a) and (b)), all tree ensembles achieve near-optimal performance, with precision and recall closely aligned across models. This result indicates that the engineered behavioral features effectively capture the consistent and often automated transaction patterns associated with large-scale cybercrime and hacking activity.

In contrast, classification performance for the \textit{Blocklisted} class (Panel (c)) remains substantially more challenging. Although ensemble methods continue to outperform the linear baseline, with F1 improvements exceeding $0.13$, this category exhibits the highest variance in detection rates. This reflects the heterogeneous nature of restricted wallet behavior, arising from post-enforcement dynamics rather than uniform illicit activity. Notably, the DNN and GNN models exhibit significant instability in this setting, particularly in recall, underscoring their sensitivity to sparse relational structures. This reinforces our conclusion that tree ensembles provide the most robust balance between precision and recall for detecting rare, heterogeneous risk events.

\subsection{Feature Importance}

To ensure the robustness of our findings and avoid the biases inherent in any single interpretability method, we implement a multi-stage Consensus Ranking pipeline. This approach identifies the most stable predictors of illicit activity by aggregating signals from three complementary importance measures:

\begin{enumerate}
    \item \textbf{Model-Specific Importance (Built-in):} We extract the native importance scores from each model. For tree ensembles, this corresponds to \textit{Gini Importance} or \textit{Mean Information Gain}, measuring how much each feature contributes to reducing node impurity across all trees. For LR, we utilize the absolute values of the normalized coefficients ($\beta$).
    
    \item \textbf{Model-Agnostic Permutation Importance:} To assess the functional impact of features on unseen data, we perform permutation importance on the full holdout test set. This method calculates the drop in the model's \textit{Macro-F1 score} when feature values are randomly shuffled, effectively measuring the model's dependence on that specific signal. We conducted 10 shuffling repetitions per feature.
    
    \item \textbf{SHAP Values:} For the ensemble models, we compute SHAP \footnote{SHAP library available at \url{https://github.com/shap}} values using the \texttt{TreeExplainer} framework to capture complex non-linear interactions. Due to the high computational cost of the Shapley value estimation, we utilized a representative subsample of 3,000 observations from the test set to bound memory requirements while maintaining a confidence interval for the importance estimates.
\end{enumerate}

For each model–importance method pair, features were sorted in descending order of importance and assigned ordinal ranks (1 = most important). For each feature, we computed the simple average of its ranks across all models and methods. Finally, features were re-ranked based on this average to obtain the Consensus Rank.

\begin{figure}[H]
    \centering
    \includegraphics[trim={3cm 2cm 3cm 1cm}, clip,angle=-90,width=1.03\linewidth]{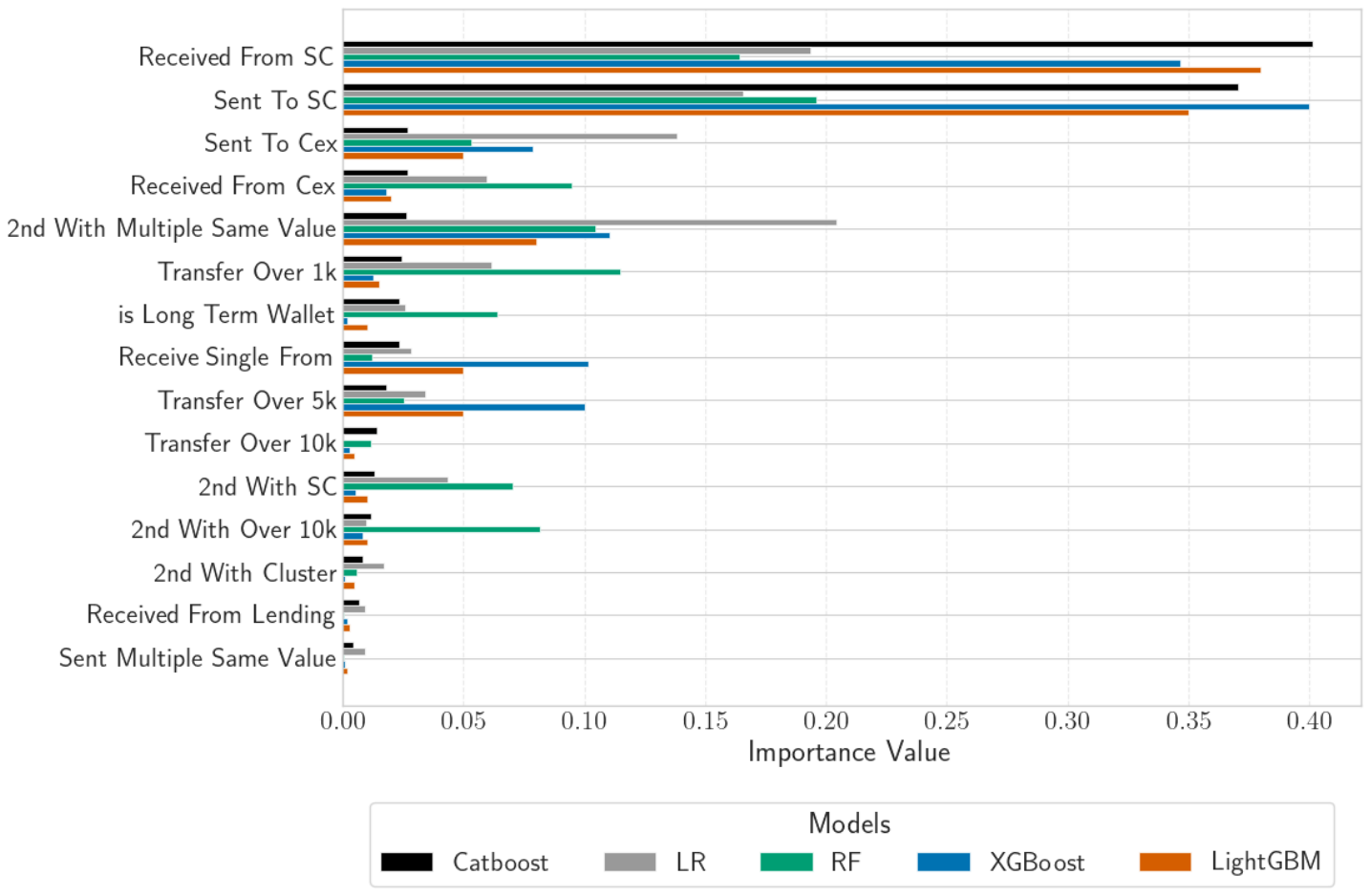} 
    \caption{Consensus ranking of the top 15 features across primary models. The reported ``Value'' represents the normalized importance derived from the multi-stage pipeline, aggregating: (i) model-specific impurity/gain for tree ensembles, (ii) normalized coefficients ($\beta$) for LR, (iii) model-agnostic permutation importance on the test set, and (iv) SHAP values. This rank-based aggregation ensures that the hierarchy reflects a stable consensus across both linear and non-linear estimators.}
    \label{Fig:top_15_importance}
\end{figure}

The results, consolidated across all models evaluated, show a clear prioritization of structural and volumetric indicators. The most influential features involve direct interactions with smart contracts (\texttt{receivedFromSC}, \texttt{sentToSC}) and volume thresholds, specifically \texttt{transferOver1k} and \texttt{transferOver5k}, which act as immediate behavioral discriminators. Furthermore, the high ranking of second-degree exposure features, such as \texttt{2ndWithMultipleSameValue} (exposure to wallets performing repeated identical-value transfers) and \texttt{2ndWithOver10k} (indirect connections to high-value flows), underscores the models' ability to look beyond immediate neighbors and identify the structural ``hop'' patterns characteristic of layering stages in money laundering.

\subsection{Class-Specific Behavioral Signatures}
\label{subsec:class_signatures}

\begin{figure}
    \centering
    \includegraphics[trim={0.5cm 3.1cm 1.6cm 3.45cm}, clip,width=1\linewidth]{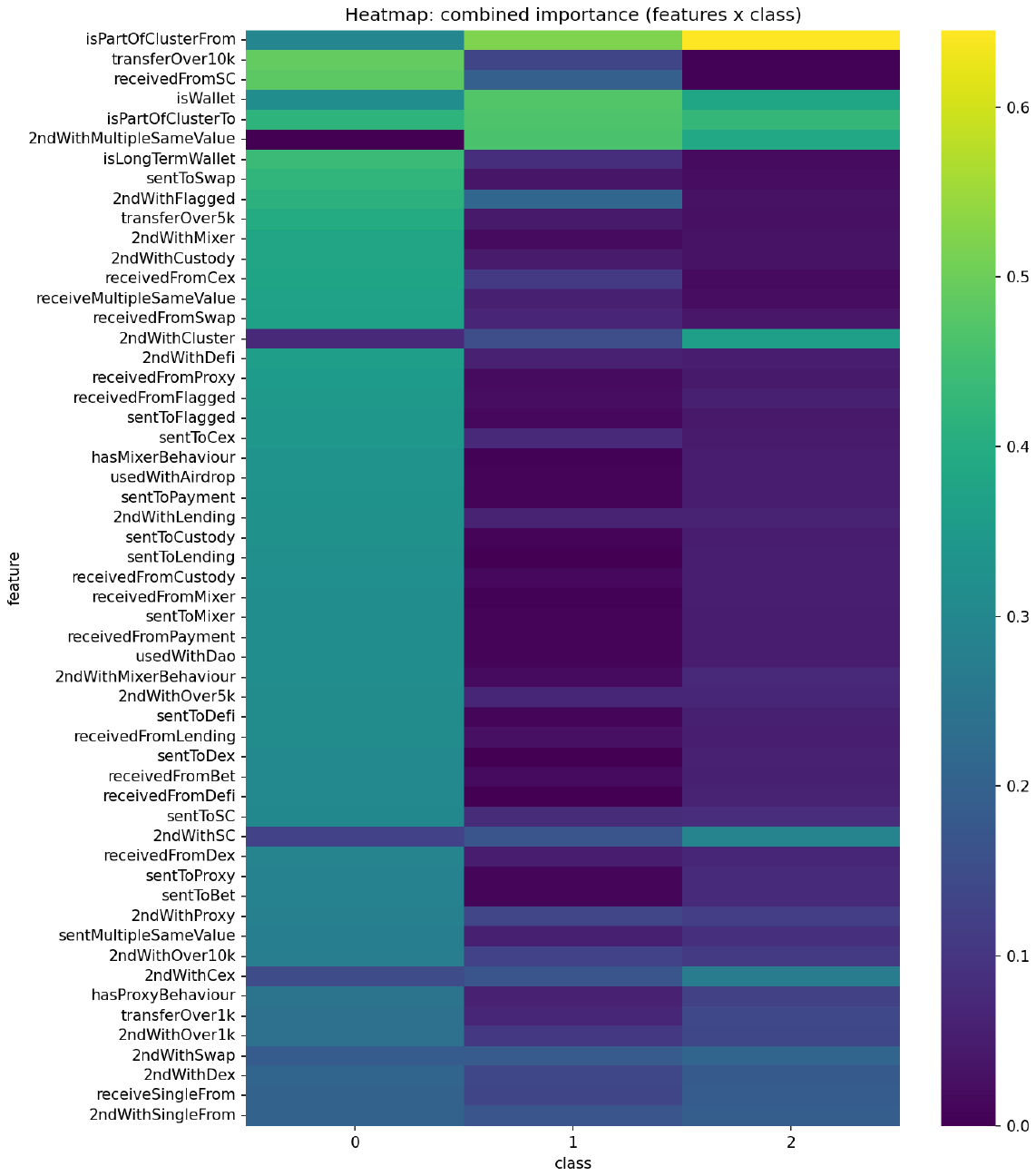}
    \caption{Class-specific feature importance heatmap. Color intensity represents the relative influence (normalized aggregated importance) of each feature on model predictions, segmented by output class. Values are averaged across all tree ensembles to highlight the distinct behavioral signatures of classes.}
    \label{fig:heatmap_importance}
\end{figure}

To examine how predictive signals differ across wallet categories, we analyze feature importance segmented by output class. Figure~\ref{fig:heatmap_importance} presents a heatmap where color intensity reflects the relative influence of each feature on class-specific predictions. 

The analysis reveals a clear behavioral separation between the risk categories. Comparing across the classes (horizontally), we observe substantial consistency: features predictive for the \textit{Normal} class generally retain relevance for the \textit{Blocklisted} class, and vice versa. However, a vertical comparison of the features within each class reveals distinct behavioral profiles that align with classical money laundering stages \citep{lin2023understanding}:

\begin{enumerate}
    \item \textit{\textbf{Normal}}: As expected, this class exhibits uniformly low feature importance across the illicit-specific indicators. The model distinguishes benign activity by the absence of the strong, coordinated signals present in the suspicious categories. Moreover, rather than relying on this absence alone, the model actively identifies benign complexity. For instance, legitimate high-frequency actors (e.g., arbitrage bots) are distinguished by their interactions with public protocols (\texttt{sentToSC}) and account longevity (\texttt{activeMonths}), rather than generic transaction volume. By contrasting these benign markers with the structural obfuscation signatures of the illicit classes (\texttt{hasProxyBehaviour}, \texttt{2ndWithFlagged}), the framework avoids penalizing standard Web3 financial complexity, ensuring it targets genuine money-laundering typologies instead.

    \item \textit{\textbf{Cybercrime}}: This class is characterized by complex network features, including cluster membership derivatives (\texttt{isPartOfClusterFrom}, \texttt{isPartOfClusterTo}) and indirect relational exposure (\texttt{2ndWithMultipleSameValue}). This deep, structured signal is consistent with the automated \textit{layering} stage. Professionalized groups route funds through numerous intermediary wallets to obscure origins, often responding to enforcement by increasing transaction hops and shifting to decentralized bridges \citep{griffin2025cryptoaml}. 

    \item \textit{\textbf{Blocklisted}}: In contrast, this class displays a concentrated reliance on direct interactions (\texttt{sentToSC}, \texttt{receivedFromCex}) and simpler network connections. This suggests behavior less driven by complex layering and more reflective of \textit{placement} or constrained post-enforcement activity. As noted in \citep{griffin2025cryptoaml}, sanctioned wallets typically exhibit reduced mobility and simpler transactional paths, often attempting reactive reallocations to non-freezable tokens like DAI.
\end{enumerate}

\begin{table}[H]
\centering
\small
\begin{tabular}{l l p{6cm}}
\hline
\textbf{Feature Group} & \textbf{AML Stage} & \textbf{Observed Behavioral Pattern} \\
\hline
SC \& CEX Flows & Placement & Direct interaction with smart contracts and centralized endpoints for initial fund entry. \\
Clustering \& Hops & Layering & High relevance of \texttt{2ndWithMultipleSameValue} and multi-hop flows for obfuscation. \\
Long-Term Retention & Integration & Use of \texttt{isLongTermWallet} and large-value transfers to identify re-integration points. \\
\hline
\end{tabular}
\caption{Principal features mapped to money-laundering typologies.}
\label{table_principal_features}
\end{table}

Beyond predictive performance, these results show that the models recover patterns closely aligned with the classical stages of money laundering:
\begin{itemize}
    \item \textbf{Placement:} Funds enter via CEXs or initial SC interactions (\texttt{receivedFromCex}, \texttt{sentToSC}), where mixers often weaken the direct link to the origin \citep{lin2023understanding}.
    \item \textbf{Layering:} Obfuscation is captured by \texttt{transferOver1k/5k/10k} and \texttt{2ndWithMultipleSameValue}, reflecting value fragmentation and proxy behavior through pass-through wallets \citep{lin2023understanding}. These signals align with the use of cross-chain bridges and DEXs that lack KYC to facilitate asset swaps.
    \item \textbf{Integration:} Re-integration is indicated by \texttt{isLongTermWallet}, identifying wallets that retain funds before redistribution, and continued mixer interactions for re-anonymization \citep{pocher2023detecting}.
\end{itemize}

To identify the limitations of the proposed framework, we conducted a qualitative analysis of the instances where the best-performing model (CatBoost) produced errors. Residual False Positives stem from legitimate high-frequency traders and arbitrage bots. These actors exhibit rapid, high-volume capital flows through centralized exchanges (CEXs) and DeFi protocols, behavioral signatures that structurally mimic the \textit{Placement} and \textit{Layering} stages of money laundering. In the multiclass setting, residual confusion arises between the \textit{Cybercrime} and \textit{Blocklisted} categories. Since both groups operate under adversarial constraints, they frequently employ overlapping obfuscation mechanisms (e.g., CEX pass-throughs and rapid swaps) prior to specific enforcement actions, creating a degree of transactional ambiguity that challenges even non-linear decision boundaries.

While these patterns strongly align with the classical stages of money laundering, we emphasize that these features output an AML risk score based on structural anomalies. They indicate a high probability of layering or obfuscation, which serves as a powerful proxy for illicit behavior, but should not be conflated with stricter legal money laundering identification without off-chain corroboration.

\subsection{Legitimate Complexity vs. Illicit Obfuscation} \label{sec:case_study}

Distinguishing legitimate high-frequency trading and arbitrage bots from laundering-related addresses represents a primary challenge in blockchain AML. These benign actors often share behavioral patterns with illicit entities, including rapid fund movement and complex protocol interactions. To clarify whether our model captures laundering-specific behavior rather than general transaction complexity, Table \ref{tab:case_study_bots} presents a feature-level comparison of three CatBoost false-positive (misclassified) arbitrage-bot wallets against three confirmed laundering addresses from the test set.

\begin{table}[H]
  \centering
  
  \begin{tabular}{l c c}
  \hline
  \textbf{Feature} & \textbf{Arbitrage Bot Examples (FP)} & \textbf{Laundering Examples} \\
  \hline
  \texttt{sentToSC} & 1 / 0 / 1 & 0  / 0 / 1 \\
  \texttt{activeMonths} & 2.1 / 1.6 / 1.9 & 1.5 / 0.1 / 0.0 \\
  \texttt{isPartOfClusterFrom} & 0 / 0 / 0 & 0 / 0 / 8 \\
  \texttt{2ndWithFlagged} & 0 / 0 / 0 & 562 / 421 / 0 \\
  \texttt{sentMultipleSameValue} & 0  / 0 / 0 & 2 / 0 / 2 \\
  \texttt{sentToCex} & 0 / 0 / 0 & 4  / 0 / 0 \\
  \texttt{receivedFromCex} & 6  / 28 / 4 & 6  / 1 / 0 \\
  \texttt{receivedFromSwap} & 3 / 0 / 1 & 0 / 0 / 2 \\
  \texttt{2ndWithSwap} & 0 /  0 / 0 & 11 / 1 / 1 \\
  \texttt{2ndWithProxy} & 0 / 0 / 0 & 0 / 0 / 30 \\
  \texttt{hasProxyBehaviour} & 0 / 0 / 0  & 5 / 3 / 3 \\
  \texttt{transferOver10k} & 0 / 0 / 0 & 11 / 9 / 16 \\
  \hline
  \end{tabular}
  \caption{Feature-level comparison of three CatBoost false-positive arbitrage-bot wallets and three confirmed laundering-related addresses. Wallet identifiers are anonymized; raw values represent a sample of top-ranking features from the test-set.}
  \label{tab:case_study_bots}
  
\end{table}

To corroborate these findings beyond the illustrative sample, Table \ref{tab:case_study_bots_average} extends the analysis to all false-positive arbitrage-bot wallets identified in the test set, reporting feature averages across the complete group. The aggregate pattern confirms the structural divergence observed in the individual examples: arbitrage-bot wallets consistently record zero values across all layering markers, while maintaining positive averages for CEX-centric features (sentToCex = 2.48, receivedFromCex = 3.07) and longer account longevity (activeMonths = 0.89). In contrast, the laundering wallets exhibit substantially elevated values on structural obfuscation indicators, including second-degree exposure to flagged entities (2ndWithFlagged = 23.39), indirect swap connectivity (2ndWithSwap = 43.80), and large-value transfer accumulation (transferOver10k = 338.29), while recording near-zero values on the CEX-interaction features that characterize the false-positive group. This population-level evidence corroborates that the residual misclassifications are driven by shared volumetric features rather than a failure of the framework to distinguish the structural signatures of money laundering.

\begin{table}[H]
  \centering
  \begin{tabular}{l c c}
  \hline
  \textbf{Feature} & Arbitrage Bot (Average) & Laundering (Average) \\
  \hline
  \texttt{sentToSC} & 0 & 0 \\
  \texttt{activeMonths} & 0.89 & 0.03 \\
  \texttt{isPartOfClusterFrom} & 0 & 0 \\
  \texttt{2ndWithFlagged} & 0 & 23.39 \\
  \texttt{sentMultipleSameValue} & 0 & 0 \\
  \texttt{sentToCex} & 2.48 & 0 \\
  \texttt{receivedFromCex} & 3.07 & 0 \\
  \texttt{receivedFromSwap} & 1.13 & 0 \\
  \texttt{2ndWithSwap} & 0 & 43.80 \\
  \texttt{2ndWithProxy} & 0 & 3.57 \\
  \texttt{2ndWithMixer} & 0 & 0.01 \\
  \texttt{hasProxyBehaviour} & 0 & 1.05 \\
  \texttt{transferOver10k} & 0 & 338.29 \\
  \hline
  \end{tabular}
  \caption{Feature-level comparison of average values over all false-positive arbitrage-bot wallets identified in the test set (N=9)}
  \label{tab:case_study_bots_average}
\end{table}

The comparison demonstrates that benign complexity and illicit obfuscation diverge along structural rather than volumetric dimensions. The arbitrage-bot wallets exhibit longer account longevity (\texttt{activeMonths} = 1.6--2.1) and CEX-centric funding, but record zero values on layering markers. The laundering wallets concentrate signal precisely on these structural features, with non-zero proxy behavior in all three cases and significant volumes of large transfers (\texttt{transferOver10k}). This evidence confirms that the model targets layering-stage topological signatures (e.g., multi-hop proxy patterns, indirect swap exposure) rather than penalizing raw transaction volume. The residual misclassifications (False Positives) stem from shared volumetric overlaps, validating that the framework fundamentally relies on AML-specific risk signals.

\subsection{\secreview{Ablation study}}
\secreview{To ensure that the performance of our models is not overly reliant on the aforementioned proximity signals, we conducted an ablation study. Table~\ref{tab:ablation_proximity} reports the performance of a representative model from each family, evaluated under the labeled-scope configuration of our main results, both with and without these known-bad-proximity screening signals (e.g., sentToFlagged, receivedFromFlagged).}

\begin{table}[H]
\centering
\begin{tabular}{llccc}
\toprule
\textbf{Model} & \textbf{Metric} & \textbf{With proximity} & \textbf{Without} & $\Delta$ \\
\midrule
CatBoost & Macro-F1 & 0.9775 & 0.9678 & $-0.0097$ \\
 & Accuracy & 0.9857 & 0.9802 & $-0.0055$ \\
 & AUROC & 0.9974 & 0.9974 & $0.0000$ \\
Random Forest & Macro-F1 & 0.9714 & 0.9690 & $-0.0024$ \\
 & Accuracy & 0.9824 & 0.9808 & $-0.0016$ \\
Logistic Regression & Macro-F1 & 0.7831 & 0.7818 & $-0.0013$ \\
DNN & Macro-F1 & 0.8708 & 0.8656 & $-0.0052$ \\
EvolveGCN-O & Macro-F1 & 0.8608 & 0.8411 & $-0.0197$ \\
 & Accuracy & 0.9107 & 0.9032 & $-0.0075$ \\
 & Recall & 0.8389 & 0.8103 & $-0.0286$ \\
 & AUROC & 0.9796 & 0.9796 & $0.0000$ \\
\bottomrule
\end{tabular}
\caption{Performance with and without label-proximity features}
\label{tab:ablation_proximity}
\end{table}

\secreview{As observed, the ablation of these features results in only a marginal decrease in overall performance metrics for the tabular models (e.g., a negligible drop in the Macro-F1 score for the CatBoost model). The GNN family exhibits a comparatively larger relative drop (EvolveGCN-O: $-0.0197$ Macro-F1, $-0.0286$ Recall), consistent with its reliance on these features as a compact substitute for structural context its message passing cannot otherwise reconstruct under the labeled-scope setting of Section~4.4. This confirms that while the proximity signals effectively emulate real-world compliance screening, the models' predictive power is predominantly driven by the broader behavioral and structural transaction patterns captured by the remaining feature set.}

\subsection{Binary Classification Results}

To benchmark the proposed framework against established baselines in the literature, we evaluate a simplified binary classification setting (Normal vs. Suspicious). In this formulation, the \textit{Cybercrime} and \textit{Blocklisted} classes are merged into a single \textit{Suspicious} category. 

To provide context, we compare our results with methods applied to the Elliptic dataset \citep{weber2019anti,pocher2023detecting}, the standard benchmark for illicit activity detection in Bitcoin. It is important to note the structural differences: while the Elliptic dataset models the Bitcoin graph, our work focuses on the Ethereum account-based model restricted to stablecoin flows. Despite these differences, both tasks share the fundamental objective of separating licit from illicit entities within high-noise transaction networks, making the comparison of methodological effectiveness relevant.

Under this reduced task, the models achieve near-perfect discrimination, confirming the strong expressiveness of the engineered feature set when the problem is simplified to separating benign from suspicious behavior. As reported in Table~\ref{table_binary_comparative}, the best-performing tree ensemble model attains an $\text{F1}$ score of $0.999$, substantially exceeding benchmarks reported in earlier work on graph-based detection. For example, studies based on the Elliptic dataset \citep{pocher2023detecting} document persistent trade-offs between precision and recall, even for advanced subgraph-based classifiers such as RevClassify, a subgraph-based deep learning model designed to mine local structural patterns to overcome label scarcity. The superior performance observed here suggests that for account-based stablecoin networks, domain-specific feature engineering (e.g., second-degree exposure and cluster derivatives) offers a marked advantage over purely topological subgraph mining. \secreview{We interpret these near-perfect binary figures, and their margin over the benchmarks above, with caution. They are obtained on a modeling dataset built through the stratified sampling and filtering funnel of Section~\ref{Sec:dataset}, which enforces a managed class balance and a proximity-cleaned Normal class, so they characterize the separability of the constructed behavioral typologies rather than the performance to be expected under the real-world class imbalance, where illicit addresses fall below 1\% of active wallets, or under the noisier labels of an operational feed. Section~\ref{Sec:conclusion} discusses this caution together with the constraints of the negative-class construction.}

\begin{table}[H]
\centering
\begin{tabular}{lcccc}
\hline
\textbf{Model} & \textbf{AUROC} & \textbf{Accuracy} & \textbf{F1-Score} & \textbf{Recall} \\
\hline
GCN (Elliptic1)                   & --     & 0.8440 & 0.9730 & -- \\
GAT (Elliptic1)                   & --     & 0.7230 & 0.9710 & -- \\
Random Forest (Elliptic1)         & --     & 0.7820 & 0.9770 & -- \\
RevClassifyDS (Elliptic2)         & 0.9740 & --     & 0.9530 & -- \\
\hline
Logistic Regression               & 0.9997 & 0.9953 & 0.9918 & 0.9959 \\
Random Forest                     & 1.0000 & 0.9997 & 0.9994 & 0.9988 \\
CatBoost                          & 1.0000 & 0.9995 & 0.9991 & 1.0000 \\
LightGBM                          & 1.0000 & 0.9998 & 0.9997 & 1.0000 \\
XGBoost                           & 1.0000 & 0.9993 & 0.9988 & 0.9994 \\
DNN                               & 0.9994 & 0.9998 & 0.9997 & 0.9994 \\
GraphSage                               & 0.9974 & 0.9929 & 0.9886 & 0.9897 \\
EvolveGCN-O                               & 0.9999 & 0.9970 & 0.9965 & 0.9962 \\
\hline
\end{tabular}
\caption{Binary classification performance (Normal vs. Suspicious) comparing prior benchmark results from the Elliptic dataset with models evaluated in this study.}
\label{table_binary_comparative}
\end{table}

By successfully mapping algorithmic importance to established financial crime typologies, the framework demonstrates that domain-informed feature engineering provides a foundation for explainable AML detection, bridging the gap between machine learning performance and regulatory interpretability.

\section{Conclusion}\label{Sec:conclusion}

The evolution of blockchain privacy presents a diverging path for compliance. While decentralized protocols increasingly adopt ZKPs to obfuscate transaction graphs, centralized stablecoins like USDT and USDC face a different reality: to maintain fiat convertibility and acceptance by regulated entities, they must guarantee auditable provenance. This creates a persistent reliance on transparent ledger histories, not merely as a technical feature, but as a prerequisite for verifying legitimate origin in an increasingly scrutinized ecosystem.

As the industry shifts from probabilistic risk assessment toward deterministic verification, where users must mathematically prove that funds are free from contact with blocklisted addresses, stablecoins will remain the critical ``transparent choke points'' of the crypto economy. This study demonstrates that analyzing only these transparent footprints allows for the reliable detection of illicit actors, even as they attempt to exit through privacy-preserving protocols.

Our benchmarking revealed that domain-informed feature engineering is the decisive factor in this setting. Tree ensembles consistently outperformed deep learning approaches, achieving Macro-F1 scores above 0.97 by effectively partitioning the feature space along discrete behavioral signals. We also successfully dissect distinct typologies, differentiating the complex, multi-hop layering of \textit{Cybercrime} syndicates from the constrained movements of \textit{Blocklisted} entities.

By mapping these algorithmic predictions to established financial crime stages, placement, layering, and integration, we establish a methodological blueprint for stablecoin AML detection. Ultimately, this scalable approach to classifying behavioral risk serves as a durable analytical baseline for compliance in an increasingly private ledger economy. 

We acknowledge several limitations of the present study. The proposed framework operates on a static, tabular feature representation and does not implement a dynamic monitoring pipeline, an adaptive labeling mechanism, or an operational deployment layer. Consequently, the mapping between predicted risk classes and the placement, layering, and integration stages of money laundering is analytical rather than prescriptive, and should be interpreted as evidence of the typologies that the engineered features can recover, rather than as a substitute for compliance decision-making. \secreview{The construction of the negative class further bounds how these results should be read. The Normal class is operationally benign under the available intelligence and filters rather than definitively benign, so some undetected illicit wallets may persist within it, and the Proximity Exclusion Filter, by removing the benign candidates nearest the illicit frontier, produces a cleaner separation between licit and illicit behavior than a deployed system would face. The near-perfect binary classification results therefore reflect the separability of the constructed dataset rather than the accuracy to be expected under real-world imbalance and label noise. These constraints temper how the numbers are interpreted, yet they leave the study's comparative finding intact, since every model is trained and evaluated under identical data conditions.} Future research could extend this baseline in several directions, integrating streaming transaction feeds to enable real-time scoring, designing adaptive labeling routines that update as new intelligence becomes available, formally validating the framework's outputs against the audit and reporting requirements of MiCA and the GENIUS Act\secreview{, and confirming external validity on temporal and incident- or cluster-level partitions and under realistic class imbalance assessed with rank-based metrics such as PR-AUC and Precision@K, before the framework is put to operational use}.

\bibliographystyle{apalike}
\bibliography{biblio}

\newpage
\appendix 

\section{Full Feature List}\label{App:features}
This appendix provides the complete list of 68 engineered features, categorized by their extraction logic and data type.

\begin{longtable}{p{3.6cm} p{7.5cm} p{0.9cm} p{1.6cm}}
\hline
\textbf{Feature Name} & \textbf{Description} & \textbf{Type} & \textbf{Source} \\
\hline
\endfirsthead
\multicolumn{4}{c}%
{ \  } \\
\hline
\textbf{Feature Name} & \textbf{Description} & \textbf{Type} & \textbf{Source} \\
\hline
\endhead
\hline \multicolumn{4}{r}{} \\
\endfoot
\endlastfoot

hasKYC & Interacted with a protocol requiring KYC & Bool. & Interaction \\
receivedFromPayment & Received funds from a payment service & Num. & Interaction \\
receivedFromBet & Received funds from a betting protocol & Num. & Interaction \\
receivedFromCex & Received funds from a centralized exchange & Num. & Interaction \\
receivedFromCustody & Received funds from a cold wallet or treasury & Num. & Interaction \\
receivedFromDefi & Received funds from a DeFi protocol & Num. & Interaction \\
receivedFromDex & Received funds from a decentralized exchange & Num. & Interaction \\
receivedFromFlagged & Received funds from a flagged address & Num. & Interaction \\
receivedFromLending & Received funds from a lending protocol & Num. & Interaction \\
receivedFromMixer & Received funds from a mixer service & Num. & Interaction \\
receivedFromSC & Received funds from a smart contract & Num. & Interaction \\
receivedFromStake & Received funds from a staking protocol & Num. & Interaction \\
receivedFromSwap & Received funds from a swap protocol & Num. & Interaction \\
sentToBet & Sent funds to a betting protocol & Num. & Interaction \\
sentToCex & Sent funds to a centralized exchange & Num. & Interaction \\
sentToCustody & Sent funds to a cold wallet or treasury & Num. & Interaction \\
sentToDefi & Sent funds to a DeFi protocol & Num. & Interaction \\
sentToDex & Sent funds to a decentralized exchange & Num. & Interaction \\
sentToFlagged & Sent funds to a flagged address & Num. & Interaction \\
sentToLending & Sent funds to a lending protocol & Num. & Interaction \\
sentToMixer & Sent funds to a mixer service & Num. & Interaction \\
sentToPayment & Sent funds to a payment service & Num. & Interaction \\
sentToSC & Sent funds to a smart contract & Num. & Interaction \\
sentToStake & Sent funds to a staking protocol & Num. & Interaction \\
sentToSwap & Sent funds to a swap protocol & Num. & Interaction \\
usedWithAirdrop & Interacted with an airdrop participant & Num. & Interaction \\
usedWithDao & Interacted with a DAO address & Num. & Interaction \\

2ndWithMixBehaviour & Second-degree connection with mixer behavior & Num. & Derived \\
2ndWithMultipleSameValue & $2^{\text{nd}}$ degree connection with repetitive same-value & Num. & Derived \\
2ndWithBet & Second-degree connection with betting wallet & Num. & Derived \\
2ndWithCex & Second-degree connection with centralized exchange & Num. & Derived \\
2ndWithCluster & Second-degree connection with clustered wallets & Num. & Derived \\
2ndWithCustody & Second-degree connection with cold wallet & Num. & Derived \\
2ndWithDefi & Second-degree connection with DeFi protocol & Num. & Derived \\
2ndWithDex & Second-degree connection with decentralized exchange & Num. & Derived \\
2ndWithFlagged & Second-degree connection with flagged address & Num. & Derived \\
2ndWithLending & Second-degree connection with lending protocol & Num. & Derived \\
2ndWithMixer & Second-degree connection with mixer & Num. & Derived \\
2ndWithOver1k & Second-degree connection with >1k transfer & Num. & Derived \\
2ndWithOver5k & Second-degree connection with >5k transfer & Num. & Derived \\
2ndWithOver10k & Second-degree connection with >10k transfer & Num. & Derived \\
2ndWithPayment & Second-degree connection with payment wallet & Num. & Derived \\
2ndWithProxy & Second-degree connection with proxy wallet & Num. & Derived \\
2ndWithSC & Second-degree connection with smart contract & Num. & Derived \\
2ndWithSingleFrom & Second-degree with single outgoing wallet & Num. & Derived \\
2ndWithSingleTo & Second-degree with single receiving wallet & Num. & Derived \\
2ndWithStaking & Second-degree connection with staking protocol & Num. & Derived \\
2ndWithSwap & Second-degree connection with swap protocol & Num. & Derived \\
3rdWithFlagged & Third-degree connection with flagged wallet & Num. & Derived \\
circleDetected & Reciprocal transfers within 24 hours & Num. & Derived \\
clusterScore &  \secreview{Participation in closed send/receive  groups} & Num. & Derived \\
hasMixerBehaviour & Imbalance between sent and received volumes & Num. & Derived \\
hasProxyBehaviour & Received and sent same amount in 24h & Num. & Derived \\
isPartOfClusterFrom & Sends funds with same value and destination & Bool. & Derived \\
isPartOfClusterTo & Receives funds with same value and origin & Bool. & Derived \\
receivedFromProxy & Received funds from proxy wallet & Num. & Derived \\
sentToProxy & Sent funds to proxy wallet & Num. & Derived \\

receiveMulSameValue & Received multiple transfers with the same value & Num. & Transfer \\
receiveSingleFrom & Multiple transfers from a single address & Num. & Transfer \\
sentMultipleSameValue & Sent multiple transfers with the same value & Num. & Transfer \\
sentToSingleAddress & Multiple transfers to a single address & Num. & Transfer \\
transferOver1k & Transfer over 1,000 USD-equivalent & Num. & Transfer \\
transferOver5k & Transfer over 5,000 USD-equivalent & Num. & Transfer \\
transferOver10k & Transfer over 10,000 USD-equivalent & Num. & Transfer \\

highFrequency & More than 10 transfers in the same day & Num. & Temporal \\
isLongTermWallet & Active across more than 3 months & Bool. & Temporal \\
isVerifiedContract & Smart contract with verified source code & Bool. & Direct \\
isWallet & Address has no bytecode (EOA) & Bool. & Direct \\

\hline
\caption{Engineered features capturing wallet behaviors and interactions.} \\
\label{tab:full_feature_list}
\end{longtable}

\pagebreak

\section{Hyperparameter Optimization}
\label{sec:hyperparams}

To ensure fair comparison and reproducibility, we performed a full \textit{Grid Search} with 5-fold cross-validation on the training set to identify the optimal hyperparameters for each model. The optimization objective was to maximize the Macro-F1 score, ensuring a balanced trade-off between precision and recall across the imbalanced classes.

For LR, we optimized the regularization strength ($C$) and penalty type using the \texttt{saga} solver, which was selected a priori to support efficient convergence on large datasets with $L_1$ penalties. For tree ensembles, which include both bagging (RF) and boosting (XGBoost, LightGBM, CatBoost) techniques, we targeted structural parameters governing tree complexity (depth, splits) and ensemble size (estimators, learning rate).

Tables~\ref{tab:logreg_params}, \ref{tab:rf_params}, and \ref{tab:boost_params} detail the search spaces defined for the grid, the rationale behind each parameter, and the final optimal values selected for the reported results.

\begin{table}[H]
\centering
\setlength{\tabcolsep}{2pt}
\renewcommand{\arraystretch}{1.2}
\begin{tabularx}{\linewidth}{l l l X}
\toprule
\textbf{Parameter} & \textbf{Search Grid} & \textbf{Best Value} & \textbf{Design Rationale} \\
\midrule
$C$ (Inverse Reg.) & $[10^{-3}, 10^{1}]$ & $10$ & \small Controls model complexity ($1/C$); higher values reduce regularization. \\
Penalty & $\{L_1, L_2\}$ & $L_1$ & \small $L_1$ promotes feature sparsity; $L_2$ prevents large weights. \\
\bottomrule
\end{tabularx}
\caption{Logistic Regression optimization. The \texttt{saga} solver was fixed for all experiments.}
\label{tab:logreg_params}
\end{table}

\begin{table}[H]
\centering

\setlength{\tabcolsep}{2pt}
\renewcommand{\arraystretch}{1.2}
\begin{tabularx}{\linewidth}{l l l X}
\toprule
\textbf{Parameter} & \textbf{Search Grid} & \textbf{Best Value} & \textbf{Design Rationale} \\
\midrule
$n_{\texttt{estimators}}$ & $[100, 500]$ & $400$ & \small Number of trees; higher improve stability. \\
min\_samples\_split & $[2, 10]$ & $5$ & \small Samples required to split an internal node. \\
min\_samples\_leaf & $[2, 5]$ & $2$ & \small Samples required at a leaf node (smoothing). \\
max\_features & $\{\text{sqrt}, \text{log2}\}$ & $\text{sqrt}$ & \small Fixed to decorrelate trees in the ensemble. \\
\bottomrule
\end{tabularx}
\caption{Random Forest (Bagging) optimization settings.}
\label{tab:rf_params}
\end{table}

\begin{table}[H]
\centering
\small 
\setlength{\tabcolsep}{2pt}
\renewcommand{\arraystretch}{1.2}
\begin{tabularx}{\linewidth}{
    >{\RaggedRight\arraybackslash\hsize=0.6\hsize}X 
    >{\RaggedRight\arraybackslash\hsize=0.8\hsize}X 
    >{\RaggedRight\arraybackslash\hsize=0.8\hsize}X 
    >{\RaggedRight\arraybackslash\hsize=1.8\hsize}X
}
\toprule
\textbf{Parameter} & \textbf{Search Grid} & \textbf{Best Values} \newline {\scriptsize (XGB/LGBM/CatB)} & \textbf{Rationale} \\
\midrule
$n_{\texttt{estimators}}$ & $[100, 700]$ & $200$ / $640$ / $300$ & \small Boosting rounds; trade-off between accuracy and speed. \\
Learning Rate & $[0.01, 0.3]$ & $0.1$ / $0.01$ / $0.25$ & \small Step size; lower values require more estimators. \\
Max Depth & $[3, 10]$ & $9$ / $10$ / $7$ & \small Controls interaction complexity and overfitting. \\
Subsample & $[0.8, 1.0]$ & $0.8$ / $0.8$ / -- & \small Fraction of data used per iteration. \\
Colsample & $[0.8, 1.0]$ & $0.8$ / $0.8$ / -- & \small Fraction of features used per tree. \\
\bottomrule
\end{tabularx}
\caption{GBMs optimization settings. CatBoost uses default sampling strategies; specific depth and iteration parameters were mapped to their framework equivalents.}
\label{tab:boost_params}
\end{table}

\section{Full Three Classification Performance Result}
This appendix details the comprehensive performance metrics for the multiclass classification task. It breaks down the AUROC, precision, recall, and F1 scores for each evaluated model across the three distinct behavioral classes: \textit{Normal} (0), \textit{Cybercrime} (1), and \textit{Blocklisted} (2). This granular view highlights the superior capacity of tree ensembles to distinguish between complex laundering schemes and static blocklisted entities compared to linear baselines.

\setlength{\LTleft}{\fill}
\setlength{\LTright}{\fill}
\setlength{\tabcolsep}{5pt}

\begin{longtable}{lcccccc}
\hline
\textbf{Model} & \textbf{Class} & \textbf{AUROC} & \textbf{Precision} & \textbf{Recall} & \textbf{F1} & \textbf{Support} \\
\hline
\endfirsthead
\textbf{Model} & \textbf{Class} & \textbf{AUROC} & \textbf{Precision} & \textbf{Recall} & \textbf{F1} & \textbf{Support} \\
\hline
\endhead
Logistic Regression & 0 & 0.9453 & 0.7877 & 0.9988 & 0.8807 & 1623 \\
&1 & 0.9351 & 0.9375 & 0.7614 & 0.8403 & 1182 \\
&2 & 0.9353 & 0.8773 & 0.4896 & 0.6285 & 482 \\
\hline
Random Forest & 0 & 0.9994 & 0.9951 & 1 & 0.9975 & 1623\\
& 1 & 0.9990 & 0.9817 & 0.9768 & 0.9792 & 1182 \\
& 2 & 0.9946 & 0.9395 & 0.9355 & 0.9375 & 482\\
\hline
CatBoost & 0 & 0.9990 & 0.9927 & 1 & 0.9963 & 1623 \\
& 1 & 0.9991 & 0.9873 & 0.9831 & 0.9852 & 1182 \\
& 2 & 0.9943 & 0.9579 & 0.944 & 0.9509 & 482 \\
\hline
LightGBM & 0 & 0.9989 & 0.9927 & 1 & 0.9963 & 1623 \\
& 1 & 0.9993 & 0.9864 & 0.9805 & 0.9835 & 1182 \\
& 2 & 0.9947 & 0.9518 & 0.9419 & 0.9468 & 482 \\
\hline
XGBoost & 0 & 0.9987 & 0.9927 & 0.9994 & 0.996 & 1623 \\
& 1 & 0.9992 & 0.9889 & 0.9805 & 0.9847 & 1182 \\
& 2 & 0.9942 & 0.9501 & 0.9481 & 0.9491 & 482 \\
\hline
DNN & 0 & 0.9896 & 0.9100 & 0.9986 & 0.9522 & 1623 \\
& 1 & 0.9518 & 0.9477 & 0.9319 & 0.9390 & 1182 \\
& 2 & 0.9436 & 0.8581 & 0.6208 & 0.7211 & 482 \\
\hline
GraphSAGE (labeled) & 0 & 0.9735 & 0.7725 & 0.9601 & 0.8557 & 1623 \\
& 1 & 0.9647 & 0.9010 & 0.7366 & 0.8109 & 1182 \\
& 2 & 0.9667 & 0.9592 & 0.6131 & 0.7481 & 482 \\
\hline
\secreview{GraphSAGE (extended)} & 0 & 0.9776 & 0.8702 & 0.9943 & 0.9281 & 1623 \\
& 1 & 0.9755 & 0.9626 & 0.9056 & 0.9332 & 1182 \\
& 2 & 0.9605 & 0.8827 & 0.6365 & 0.7397 & 482 \\
\hline
EvolveGCN-O (labeled) & 0 & 0.9922 & 0.9075 & 0.9968 & 0.9501  & 1623 \\
& 1 & 0.9865 & 0.9207 & 0.9419 & 0.9312  & 1182 \\
& 2 & 0.9602 & 0.8912 & 0.5782 & 0.7014 & 482 \\
\hline
\secreview{EvolveGCN-O (extended)} & 0 & 0.9935 & 0.8619 & 0.9968 & 0.9245 & 1623 \\
& 1 & 0.9844 & 0.9631 & 0.8729 & 0.9158 & 1182 \\
& 2 & 0.9517 & 0.8469 & 0.6324 & 0.7241 & 482 \\
\hline
EvolveGCN-H  (labeled) & 0 & 0.9886 & 0.8412 & 0.9931 & 0.9108 & 1623 \\
& 1 & 0.9652 & 0.9574 & 0.8600 & 0.9061 & 1182 \\
& 2 & 0.9534 & 0.8402 & 0.5821 & 0.6877 & 482 \\
\hline
\secreview{EvolveGCN-H (extended)} & 0 & 0.9935 & 0.8943 & 0.9880 & 0.9388 & 1623 \\
& 1 & 0.9844 & 0.9433 & 0.9102 & 0.9265 & 1182 \\
& 2 & 0.9517 & 0.8337 & 0.6388 & 0.7234 & 482 \\
\hline 
\caption{Detailed multiclass performance metrics by model. The labels correspond to Normal (0), Cybercrime (1), and Blocklisted (2) entities.} \\
\label{table_three_class_results}
\end{longtable}


\end{document}